\documentclass[prd,aps,nofootinbib,notitlepage]{revtex4}
\usepackage{graphicx,epsf,amsmath,amsfonts,amssymb,amsbsy,textcomp}
\usepackage{epsfig}
\newcommand{\vev}[1]{\langle#1\rangle}
\newcommand{\mat}{\left ( \begin{array}}
\newcommand{\emat}{\end{array} \right )}
\newcommand{\vect}{\left ( \begin{array}{c}}
\newcommand{\evect}{\end{array} \right )}

\begin{document}


\title{
The dual properties of chiral and isospin asymmetric dense quark matter formed of two-color quarks

}
\author{T. G. Khunjua $^{1),~2)}$, K. G. Klimenko $^{3)}$, and R. N. Zhokhov $^{3),~4)}$ }

\affiliation{$^{1)}$ The University of Georgia, GE-0171 Tbilisi, Georgia}
\affiliation{$^{3)}$ State Research Center
of Russian Federation -- Institute for High Energy Physics,
NRC "Kurchatov Institute", 142281 Protvino, Moscow Region, Russia}
\affiliation{$^{4)}$  Pushkov Institute of Terrestrial Magnetism, Ionosphere and Radiowave Propagation (IZMIRAN),
108840 Troitsk, Moscow, Russia}

\begin{abstract}

In this paper the phase structure of dense baryon matter composed of $u$ and $d$ quarks with two colors has been investigated in the presence
of baryon $\mu_B$, isospin $\mu_I$ and chiral isospin $\mu_{I5}$  chemical potentials in the framework of Nambu--Jona-Lasinio
model with quark-antiquark and quark-quark interaction channels. In the chiral limit, it has been shown in the mean-field approximation that
the duality  between phases with spontaneous chiral symmetry breaking and condensation of charged pions, found in the three color case, remains
valid in the two color case. In addition, it has been shown that there are two more dualities in the phase diagram in two color case,
namely (as in the case with $\mu_{I5}=0$), at $\mu_{I5}\ne 0$ the general $(\mu,\mu_I,\mu_{I5})$-phase portrait of the model has dual symmetry
between the phase with condensation of charged pions and the phase with diquark condensation.
This duality stays exact even in the physical point, $m_0\ne 0$. And at $m_0=0$ the $(\mu,\mu_I,\mu_{I5})$-phase portrait becomes even more
symmetrical, since dual symmetry between phases with spontaneous chiral symmetry breaking and diquark condensation appears.
It is shown that due to the dualities the phase diagram is extremely symmetric and has interlacing structure.
One can show that 
the phase portrait of two-color NJL model can be obtained just
by duality properties from the results of investigations of three-color NJL model (it was noticed only after the numerical calculations have
been performed).
Three-color case shares only one duality of the two color one, and one can only see a facet of this enormously symmetric picture in the case
of three colors.  Using dualities only, it is possible to show that there are no mixed phases (phases with two non-zero condensates).
This prediction of dualities is of great use, because for sure it can be shown by the direct calculations
but it would be enormously more complicated and time-consuming numerically.
\end{abstract}

\maketitle

\section{Introduction}
It is well known that quantum chromodynamics (QCD) is the theoretical basis for the investigation of strongly interacting matter. However, to study the properties of dense quark (baryon) matter, which can exist in the cores of compact stars or arise in heavy ion collisions, the perturbative QCD methods are not applicable because the coupling constant of strong interactions is quite large in these cases. Another well-known (and nonperturbative) method of QCD research, the lattice approach, is also not applicable in this case due to a notorious sign problem.
Thus, interest in the phenomena occurring in dense baryonic matter serves as a trigger for studying QCD-like models and/or theories that are free from the above disadvantages and contribute to a deeper understanding of the properties of the phase diagram of matter at nonzero temperature and baryon density. Among such theories, for example, there are Nambu--Jona-Lasinio (NJL)-like models \cite{njl} (see also the reviews \cite{buballa,zhokhov} and references therein), the QCD and QCD-like models with two-color quarks \cite{kogut,son2,weise,ramos, andersen3,brauner1,andersen2,imai,adhikari,chao,Bornyakov:2020kyz}, etc.

In addition to the non-zero density of the baryon charge, quark matter can have a non-zero density of isospin (isospin asymmetry), which is an evident characteristic of neutron stars. Such a phenomenon as the charged pion condensation (PC) is closely related to the isotopic
asymmetry of matter. And the question of whether the charged PC phase can exist in dense quark matter has been discussed \cite{son, he,abuki,ak,ekkz,Mammarella:2015pxa,Andersen:2018nzq}. Note that usually dense baryonic matter with isospin imbalance is described in terms of baryon $\mu_B$- and
isospin $\mu_I$ chemical potentials. 

Quite recently, it was understood that chiral asymmetry (or chiral imbalance), i.e. unequal
densities $n_L$ and $n_R$ of all left- and all right-handed quarks, is also one of the properties of dense quark matter.
Usually, chiral asymmetry is characterized by a quantity $n_5$ called  
chiral density, $n_5\equiv n_R-n_L$. It can be generated dynamically at high temperatures, for example, in the fireball after heavy ion
collision, by virtue of the Adler-Bell-Jackiw anomaly and quarks interacting with gauge (gluon) field configurations with nontrivial topology,
named sphalerons. In the presence of an external strong magnetic field, which can be produced in heavy ion collisions as well, this can lead
to the so-called chiral magnetic effect \cite{fukus}. Moreover, in the presence of external magnetic field chiral density
$n_5$ can be produced (even at rather low temperature) in dense quark matter due to the so-called chiral separation effect \cite{Metlitski}
(it can be also produced under fast rotations of the system due to the so-called chiral vortical effect). Now, let us notice that usually
when one talk about chiral density $n_5$ one implies that chiral density $n_{u5}$ of $u$ quarks and chiral density $n_{d5}$ of $d$ quarks
are equal to each other (it is evident that $n_{5}=n_{u5}+n_{d5}$). Indeed, that is the case when one has in mind the mechanism of
generation of chiral imbalance at high temperatures due to nontrivial topology of gauge field configuration. In this case it is quite
plausible that $n_{u5}=n_{d5}$ due to the fact that gluon field interacts with different quark flavors in exactly the same way and does not
feel the difference between flavors. But another mechanism, the chiral separation effect, is sensitive to the flavor of quarks
(as it was shown in Appendices A to the papers \cite{kkz2}). So in dense quark matter a strong magnetic field separates $u$ and $d$
quarks in different ways. As a result, we see that, e.g., in such astrophysical objects as magnetars there might exist areas, in which
the quantity $n_{I5}=n_{u5}-n_{d5}$, called the chiral isospin density, is not zero. Moreover, it has been argued that chiral imbalance
is generated by parallel magnetic and electric fields \cite{Ruggieri:2016fny}, one can generalize these arguments to chiral isospin imbalance
as well.

So in the most general case, chiral imbalance of quark matter is described by two chemical potentials,
chiral $\mu_5$ and chiral isospin (or isotopic) $\mu_{I5}$ chemical potentials, which are thermodynamically conjugated to
$n_{5}$ and $n_{I5}$, respectively. The first, $\mu_5$, is usually used when isospin
asymmetry of quark matter is absent, i.e. in the case $\mu_I=0$ \cite{andrianov,cao,braguta}. The second, $\mu_{I5}$, might be taken
into account when, in addition to chiral, there is also isotopic asymmetry of matter, in which charged PC phenomenon may occur, etc.
\cite{kkz}. In particular, it was established in the framework of NJL models that $\mu_5$ catalyzes the chiral symmetry
breaking (CSB) \cite{braguta,kkz}, whereas $\mu_{I5}$ promotes the charged PC in dense quark matter \cite{kkz}. It should be noted
once more that chiral asymmetry of baryonic matter can occur under the influence of a strong magnetic field due to chiral magnetic
and chiral separation effects, i.e. chiral imbalance is an inevitable characteristic of a dense baryonic (quark) medium in compact stars, as well as in the collision of heavy ions.

In our recent studies \cite{kkz2,kkz,kkz1+1}, we showed that chiral asymmetry contributes to the generation of charged PC phenomenon
in dense quark matter. Moreover, as it was demonstrated in these papers, in the presence of isospin ($\mu_I\ne 0$) and chiral isospin
asymmetry ($\mu_{I5}\ne 0$) there is a duality between CSB and charged PC phenomena. It means that there is a symmetrical arrangement of these phases on the full phase diagram of quark matter. Our recent studies were carried out in the framework of the NJL model in which the diquark interaction channel, and hence the color superconductivity (CSC) phenomenon, was not
taken into account. It is well known that CSC can appear in quark medium at densities several times higher than the density of ordinary
nuclear matter, i.e. for $\mu_B\gtrsim 1$ GeV (see, e.g., in reviews \cite{alford} devoted to the CSC phenomenon).
Therefore, despite the fact that our results are in well agreement with the
lattice QCD analysis \cite{kkz2}, we can trust the duality properties of dense quark matter only in the range of rather low values of the baryonic chemical potential $\mu_B< 1$ GeV. On the other hand, inside of neutron stars
baryon densities can reach rather large values (corresponding to higher values of $\mu_B$), which means that CSC can be present there.
Hence, it is interesting to study how the
chiral asymmetry of baryonic matter, which, as mentioned above, is one of the essential features of compact stars with a strong magnetic
field, correlates with the phenomenon of color superconductivity. Just the goal of the present paper is to shed new light on the role of both
isospin and chiral isospin asymmetries in the formation of diquark condensation, i.e. color superconductivity, at zero temperature.
At the same time, to simplify
the analysis and avoid unnecessary computational difficulties, we use an approach based on the NJL model, in which quarks are two color.

As an additional motivation for the above proposed two-color approach to quarks, it can be noted that in the QCD with two-color gauge symmetry
there is no sign problem of the Dirac operator determinant, so in the framework of this theory first principle lattice calculations can be used
to study dense quark-hadron matter (see, e.g., in Refs. \cite{kogut}). Moreover, an analysis of dense hadron matter carried out
in the two-color NJL model by the mean-field method, showed a good agreement with corresponding lattice results
\cite{weise,ramos,andersen3,brauner1,andersen2,imai,chao,Bornyakov:2020kyz}. Since in the two-color NJL model not only mesons, but also the simplest colorless
baryons are composed of quarks, we have a good opportunity to simulate
the meson-baryon interaction both in vacuum and in dense medium, relying on a uniform quark model. Furthermore, in this case the properties
of mesons, baryons as well as different phenomena of dense hadron matter, such as diquark condensation, etc, can be investigated
with the help of well-known analytical methods. In principle, these results can then be generalized (or extrapolated) to real physical situation with three color quarks, etc.

The paper is organized as follows.
In Sec. II a (3+1)-dimensional NJL model with two quark flavors ($u$ and $d$ quarks), each is a color doublet, is introduced. It
contains both quark-antiquark and diquark channels of interaction and, as it was shown in Ref. \cite{weise}, is a low energy effective model of
the two-color QCD. Then, in order to simulate the properties of real dense quark-hadron matter in the framework of this model, we introduce
three chemical potentials, baryon $\mu_B$, isospin $\mu_I$ and chiral isospin $\mu_{I5}$ and find in the mean-field approximation the
thermodynamic potential (TDP) dependent on chiral, diquark, and charged pion condensates. In Sec. III three duality properties (dual
symmetries) of the model TDP are established. Each duality property of the model means that its TDP is invariant under some interchange
of chemical potentials as well as simultaneous interchange of corresponding condensates. Using dual symmetries of the TDP, in this section
a phase structure for some simple chemical potential contents of the model is established. For example, if the phase structure of the model
is known at $\mu_B\ne 0$ and $\mu_I=\mu_{I5}=0$ \cite{weise}, then using a dual mapping it is possible to find its phase structure without
any calculations at $\mu_{I5}\ne 0$ and $\mu_B=\mu_I=0$, etc. In Sec. IV the general $(\mu_B,\mu_I,\mu_{I5})$-phase diagram
of the model is analyzed on the basis of dual symmetries of the TDP. In particular, it is shown here that the inclusion of a diquark
interaction channel in the analysis does not destroy the dual symmetry of dense quark matter between CSB and charged PC phases,
previously found in Refs. \cite{kkz2,kkz,kkz1+1}. Moreover, the new symmetries between CSB, charged PC and diquark condensation phases
of quark matter appear in the general phase portrait of the model. In Sec. V summary and conclusions are given. Some technical
details are relegated to three Appendixes.


\section{Two-color (3+1)-dimensional NJL model and its thermodynamic potential}

It is well known that in the chiral limit the two-color QCD with $u$ and $d$ quarks is symmetrical not only with respect to the usual flavor $SU(2)_L\times SU(2)_R$ group, but also with an enlarged flavor $SU(4)$ group of transformations of quark fields \cite{kogut,son2}. Since the $SU(4)$ symmetry
connects quarks and antiquarks, it is usually called Pauli-Gursey symmetry. In two-color QCD the colorless baryons are formed by two quarks, i.e. baryons in this theory are bosons.

To obtain an effective 4-fermion NJL model that would reproduce the basic low-energy properties of the initial two-color QCD theory and had the same symmetry group, the
authors of Ref. \cite{weise} integrated out the gluon fields in the generated functional of the QCD. Then, after ``approximating'' the nonperturbative gluon propagator by a $\delta-$function, one arrives at an effective local chiral four-quark interaction of the form (color current)$\times$(color current) of the NJL type describing low-energy hadron physics. Finally, by performing a Fierz transformation of this interaction term and taking into account only scalar and pseudo-scalar $(\bar q q)$- as well as scalar
$(qq)$-interaction channels, one obtains a four-fermionic model given by the following Lagrangian (in Minkowski space-time notation) \footnote{The most general Fierz transformed four-fermion
interaction includes additional vector and axial-vector $(\bar q q)$ as well as pseudo-scalar, vector and axial-vector-like $(qq)$-interactions. However, these terms are omitted here for simplicity.}
\begin{eqnarray}
L=\bar q \Big [i\hat\partial-m_0\Big ]q+H\Big [(\bar qq)^2+(\bar qi\gamma^5\vec\tau q)^2+
\big (\bar qi\gamma^5\sigma_2\tau_2q^c\big )\big (\overline{q^c}i\gamma^5\sigma_2\tau_2 q\big )\Big]. \label{IV.1}
\end{eqnarray}
In (\ref{IV.1}), $\hat\partial\equiv\gamma^\rho\partial_\rho$; 
$q^c=C\bar q^T$, $\overline{q^c}=q^T C$ are charge-conjugated spinors, and
$C=i\gamma^2\gamma^0$ is the charge conjugation matrix (the symbol
$T$ denotes the transposition operation). The quark field
$q\equiv q_{i\alpha}$ is a flavor and color doublet as
well as a four-component Dirac spinor, where $i=1,2$ or $u,d$; $\alpha =
1,2$. (Latin and Greek indices refer to flavor and color
indices, respectively; spinor indices are omitted.) Furthermore,
we use the notations  $\vec \tau\equiv (\tau_{1},
\tau_{2},\tau_3)$ and $\sigma_2$ for usual Pauli matrices acting in the two-dimensional flavor and color spaces, respectively. The Lagrangian (\ref{IV.1}) is invariant under color $SU(2)_c$ and baryon $U(1)_B$ symmetries. Moreover, at $m_0=0$, it has the same Pauli-Gursey flavor $SU(4)$ symmetry as two-color QCD.

Generally speaking, Lagrangian (\ref{IV.1}) describes physical processes in a vacuum. In order to study the physics of dense quark matter, it is necessary to add in Eq. (\ref{IV.1}) several terms with chemical potentials,
\begin{eqnarray}
L\longrightarrow L_{dense}=L+\bar q {\cal M}\gamma^0q\equiv L+\bar q \left[\frac{\mu_B}{2}+\frac{\mu_I}2\tau_3
+\frac{\mu_{I5}}2\gamma^5\tau_3\right]\gamma^0q. \label{IV.100}
\end{eqnarray}
The Lagrangian $L_{dense}$ contains baryon $\mu_B$-, isospin $\mu_I$- and chiral isospin $\mu_{I5}$ chemical potentials. In other words, this model is able to describe the properties of quark matter with nonzero baryon $n_B=(n_{u}+n_{d})/2\equiv n/2$,
isospin $n_I=(n_{u}-n_{d})/2$ and chiral isospin $n_{I5}=(n_{u5}-n_{d5})/2$ densities, which are the quantities
thermodynamically conjugated to chemical potentials $\mu_B$, $\mu_I$ and $\mu_{I5}$, respectively.
(Above we use the notations $n_f$ and $n_{fL(R)}$ for density of quarks as well as density of left(right)-handed quarks with individual flavor $f=u,d$, respectively. Moreover, $n_{f5}=n_{fR}-n_{fL}$.) Note that at $m_0=0$ the Lagrangian (\ref{IV.100}), due to the terms with chemical potentials, is no longer invariant with respect to Pauli-Gursey $SU(4)$ symmetry. So, in the chiral limit the Lagrangian (\ref{IV.100}), apart from the color $SU(2)_c$, is invariant with respect to the $U(1)_B\times U(1)_{I_3}\times U(1)_{AI_3}$ group, where
\begin{eqnarray}
U(1)_B:~q\to\exp (\mathrm{i}\alpha/2) q;~
U(1)_{I_3}:~q\to\exp (\mathrm{i}\alpha\tau_3/2) q;~
U(1)_{AI_3}:~q\to\exp (\mathrm{i}
\alpha\gamma^5\tau_3/2) q.
\label{2001}
\end{eqnarray}
Moreover, the quantities $n_B$, $n_I$ and $n_{I5}$ are the ground state expectation values of the densities of conserved charges corresponding to $U(1)_B$, $U(1)_{I_3}$ and $U(1)_{AI_3}$ symmetry groups. So we have from (\ref{2001}) that $n_B=\vev{\bar q\gamma^0q}/2$, $n_I=\vev{\bar q\gamma^0\tau^3 q}/2$ and $n_{I5}=\vev{\bar q\gamma^0\gamma^5\tau^3 q}/2$. We would like also to remark that, in addition to (\ref{2001}), the Lagrangian (\ref{IV.100}) is invariant with respect to the electromagnetic $U(1)_Q$ group,
\begin{eqnarray}
U(1)_Q:~q\to\exp (\mathrm{i}Q\alpha) q,
\label{2002}
\end{eqnarray}
where $Q={\rm diag}(q_u,q_d)$ (here $q_u$, $q_d$ are the electric charges of $u$ and $d$ quarks, respectively). 
It is natural to assume that in this theory $u$ and $d$ quarks have a baryon
charge equal to 1/2, and their electric charges are 1/2 and (-1/2), correspondingly \cite{ramos}. In this case scalar colorless diquarks with unit baryon charge have zero electric charge. Their condensation leads to a phase, which is naturally called the (baryon) superfluid phase (by analogy with the phenomenon of superfluidity of helium). At the same time, in some papers on two-color QCD,
it is assumed that the electric charge of $u$ quark is 2/3, and that of the $d$ quark is (-1/3) \cite{andersen3}. In this case colorless diquarks of the model have electric charge 1/3. However, since in our consideration both the electromagnetic interaction between quarks and strong external magnetic (electric) fields are ignored, the results of the study of the phase structure of the model do not depend on how to interpret the electric charges of quarks.

The ground state expectation values of $n_B$, $n_I$ and $n_{I5}$ 
can be found by differentiating the thermodynamic potential (TDP) of the model (\ref{IV.100}) with respect to the corresponding chemical potentials. The goal of the present paper is the investigation of the ground state properties (or phase structure) of the system (\ref{IV.100}) and its dependence on the chemical potentials  $\mu_B$, $\mu_I$ and $\mu_{I5}$.

To find the TDP, we starting from a semibosonized (linearized) version $\widetilde L$ of the Lagrangian (\ref{IV.100}) that contains auxiliary bosonic fields $\sigma (x)$, $\vec\pi  =(\pi_1 (x),\pi_2 (x),\pi_3 (x))$, $\Delta (x)$ and $\Delta^* (x)$ and has the following form
\begin{eqnarray}
\widetilde L=\bar q \Big [i\hat\partial-m_0+{\cal M}\gamma^0 -\sigma -i\gamma^5\vec\tau\vec\pi\Big ]q-\frac{\sigma^2+\vec\pi^2+
\Delta^*\Delta}{4H}-\frac{\Delta}{2}\Big [\bar qi\gamma^5\sigma_2\tau_2q^c\Big ]-
\frac{\Delta^*}{2}\Big [\overline{q^c}i\gamma^5\sigma_2\tau_2 q\Big ], \label{IV.2}
\end{eqnarray}
where ${\cal M}$ is introduced in Eq. (\ref{IV.100}). Clearly, the Lagrangians (\ref{IV.100}) and (\ref{IV.2}) are equivalent, as can be seen by using the Euler-Lagrange
equations of motion for bosonic fields which take the form
\begin{eqnarray}
\sigma (x)=-2H(\bar qq),&~&\Delta (x)= -2H\Big [\overline{q^c}i\gamma^5\sigma_2\tau_2 q\Big ]=-2H\Big [q^TCi\gamma^5\sigma_2\tau_2 q\Big ],\nonumber\\
~\vec\pi(x)=-2H(\bar qi\gamma^5\vec\tau q),&~&
\Delta^*(x)=-2H\Big [\bar qi\gamma^5\sigma_2\tau_2q^c\Big ]=-2H\Big [\bar qi\gamma^5\sigma_2\tau_2C\bar q^T\Big ].\label{IV.3}
\end{eqnarray}
It is easy to see from Eq. (\ref{IV.3}) that $\sigma(x)$ and $\pi_a(x)$ ($a=1,2,3$) are Hermitian, i.e. real, bosonic fields, whereas $\Delta^*(x)$ and $\Delta(x)$ are Hermitian conjugated to each other. Indeed, one can check that $(\sigma(x))^\dagger=\sigma(x)$,
$(\pi_a(x))^\dagger=\pi_a(x)$, $(\Delta(x))^\dagger=\Delta^*(x)$ and $(\Delta^*(x))^\dagger=\Delta(x)$, where the superscript symbol $\dagger$ denotes the Hermitian conjugation. Note that the composite bosonic field $\pi_3 (x)$ can be identified with the physical $\pi^0(x)$-meson field, whereas the physical $\pi^\pm (x)$-meson fields are the following combinations of the composite fields, $\pi^\pm (x)=(\pi_1 (x)\mp i\pi_2 (x))/\sqrt{2}$. It is clear that the groung state expectation values of all bosonic fields (\ref{IV.3})
are $SU(2)_c$ invariant, hence in this model the color symmetry can not be broken dynamically (spontaneously). If the ground state expectation values $\vev{\sigma (x)}\ne 0$ or $\vev{\pi_0 (x)}\ne 0$, then chiral symmetry $U(1)_{AI_3}$ of the model (\ref{IV.100}) is broken spontaneously.
If in the ground state we have $\vev{\pi_{1,2} (x)}\ne 0$, then both the isospin  $U(1)_{I_3}$ and the electromagnetic $U(1)_Q$ symmetries are broken spontaneously. This phase of quark matter is called the charged pion condensation (PC) phase. Finally, if $\vev{\Delta (x)}\ne 0$, then in
the system spontaneous breaking of the baryon $U(1)_B$ symmetry occurs, and the baryon superfluid (BSF) phase is realized in the model. Note that in the BSF phase the electromagnetic $U(1)_Q$ symmetry remains intact if $Q={\rm diag} (1/2,-1/2)$.

Introducing the Nambu-Gorkov bispinor field $\Psi$, where
\begin{equation}
\Psi=\left({q\atop q^c}\right),~~\Psi^T=(q^T,\bar q C^{-1});~~
\quad \overline\Psi=(\bar q,\overline{q^c})=(\bar q,q^T C)=\Psi^T \left
(\begin{array}{cc}
0~~,&  C\\
C~~, &0
\end{array}\right )\equiv\Psi^T Y,\label{IV.5}
\end{equation}
one can present the auxiliary Lagrangian (\ref{IV.2}) in the following form
\begin{eqnarray}
\widetilde L=-\frac{\sigma^2+\vec\pi^2+\Delta^*\Delta}{4H}+\frac 12\Psi^T(YZ)\Psi, \label{IV.12}
\end{eqnarray}
where matrix $Y$ is given in Eq. (\ref{IV.5}) and
\begin{equation}
Z=\left (\begin{array}{cc}
D^+, & K\\
K^*~~, &D^-
\end{array}\right )\equiv \left (\begin{array}{cc}
i\hat\partial-m_0+{\cal M}\gamma^0 -\sigma -i\gamma^5\vec\tau\vec\pi, & -i\gamma^5\sigma_2\tau_2\Delta\\
~~~~~~~~-i\gamma^5\sigma_2\tau_2\Delta^*~~~~~~~~~~~~, &i\hat\partial-m_0-\gamma^0{\cal M} -\sigma -i\gamma^5(\vec\tau)^T\vec\pi
\end{array}\right ).\label{IV.13}
\end{equation}
Notice that matrix elements of the 2$\times$2 matrix $Z$, i.e. the quantities $D^\pm$, $K$ and $K^*$, are the nontrivial operators
in the (3+1)-dimensional coordinate, four-dimensional spinor, 2-dimensional flavor and ($N_c=2$)-dimensional color spaces.
Then, in the one fermion-loop (or mean-field) approximation, the effective action ${\cal S}_{\rm
{eff}}(\sigma,\vec\pi,\Delta,\Delta^{*})$ of the model (\ref{IV.100}) (this quantity is the generating functional of a one-particle irreducible
Green functions of bosonic fields (\ref{IV.3})) is expressed by means of the path integral over quark fields:
\begin{eqnarray}
\exp(i {\cal S}_{\rm {eff}}(\sigma,\vec\pi,\Delta,\Delta^{*}))=
  N'\int[d\bar q][dq]\exp\Bigl(i\int\widetilde L\,d^4 x\Bigr),\label{IV.14}
\end{eqnarray}
where $N'$ is a normalization constant and
\begin{eqnarray}
&&{\cal S}_{\rm {eff}}
(\sigma,\vec\pi,\Delta,\Delta^{*})
=-\int d^4x\left [\frac{\sigma^2(x)+\vec\pi^2(x)+|\Delta(x)|^2}{4H}\right ]+
\widetilde {\cal S}_{\rm {eff}}.\label{IV.15}
\end{eqnarray}
The quark contribution to the effective action, i.~e.\  the term
$\widetilde {\cal S}_{\rm {eff}}$ in (\ref{IV.15}), is given by:
\begin{equation}
\exp(i\tilde {\cal S}_{\rm {eff}})=N'\int [d\bar
q][dq]\exp\Bigl(\frac{i}{2}\int\Big [\Psi^T(YZ)\Psi\Big ]d^4 x\Bigr).
\label{IV.16}
\end{equation}
Note that in Eqs. (\ref{IV.14})-(\ref{IV.16}) we have used the expression (\ref{IV.12}) for the auxiliary Lagrangian $\widetilde L$.
Since the integration measure in Eq. (\ref{IV.16}) obeys the relation $[d\bar q][dq]=$ $[d q^c][dq]=$ $[d\Psi]$, we have from it
\begin{equation}
\exp(i\tilde {\cal S}_{\rm {eff}})=
  \int[d\Psi]\exp\left\{\frac i2\int\Psi^T(YZ)\Psi
  d^4x\right\}=\mbox {det}^{1/2}(YZ)=\mbox {det}^{1/2}(Z),\label{IV.17}
\end{equation}
where the last equality is valid due to the evident relation $\det
Y=1$. Then, using the Eqs (\ref{IV.15}) and (\ref{IV.17}) one can obtain the following
expression for the effective action (\ref{IV.15}):
\begin{equation}
{\cal S}_{\rm
{eff}}(\sigma,\vec\pi,\Delta,\Delta^{*})
=-\int d^4x\left[\frac{\sigma^2(x)+\vec\pi^2(x)+|\Delta(x)|^2}{4H}\right]-\frac i2\ln\mbox {det}(Z).\label{IV.18}
\end{equation}
Starting from (\ref{IV.18}), one can define in the
mean-field approximation the thermodynamic potential (TDP) $\Omega(\sigma,\vec\pi, \Delta, \Delta^{*})$
of the model (\ref{IV.100})-(\ref{IV.2}) at zero temperature,
\begin{equation}
{\cal S}_{\rm {eff}}~\bigg
|_{~\sigma,\vec\pi,\Delta,\Delta^{*}=\rm {const}}
=-\Omega(\sigma,\vec\pi,\Delta,\Delta^{*})\int d^4x.
\label{IV.19}
\end{equation}
The ground state expectation values (mean
values) of the fields,
$\vev{\sigma(x)}\equiv\sigma,~\vev{\vec\pi(x)}
\equiv\vec\pi,~\vev {\Delta(x)}\equiv\Delta,~
\vev{\Delta^{*}(x)}\equiv\Delta$, are
solutions of the gap equations for the TDP $\Omega$ (below, in our approach all ground state expectation values
$\sigma,\vec\pi,\Delta,\Delta^*$ do not depend on coordinates $x$):
\begin{eqnarray}
\frac{\partial\Omega}{\partial\pi_a}=0,~~~~~
\frac{\partial\Omega}{\partial\sigma}=0,~~~~~
\frac{\partial\Omega}{\partial\Delta}=0,~~~~~
\frac{\partial\Omega}{\partial\Delta^{*}}=0.\label{IV.20}
\end{eqnarray}
Since the matrix $Z$ in Eq. (\ref{IV.18}) has an evident 2$\times$2 block structure (see in Eq. (\ref{IV.13})), one can use there a general formula
\begin{eqnarray}
\det\left
(\begin{array}{cc}
A~, & B\\
C~, & D
\end{array}\right )=\det [-CB+CAC^{-1}D]=\det
[DA-DBD^{-1}C],\label{IV.21}
\end{eqnarray}
and find that (taking into account the relation $\tau_2\vec\tau\tau_2=-\vec\tau^T$ and assuming that all bosonic fields do not depend on $x$)
\begin{eqnarray}
&&\det(Z)\equiv\det\left
(\begin{array}{cc}
D^+~, & K\\
K^*~, & D^-
\end{array}\right )=\det\big(-K^*K+K^*D^+K^{*-1}D^-\big)\nonumber\\
&&=\det \Big [\Delta^*\Delta+\big (-i\hat\partial-m_0-\widetilde{\cal M}\gamma^0 -\sigma +i\gamma^5(\vec\tau)^T\vec\pi\big )\big (i\hat\partial-m_0-
\gamma^0{\cal M} -\sigma -i\gamma^5(\vec\tau)^T\vec\pi\big )\Big ],\label{IV.22}
\end{eqnarray}
where (in the following we use also the notations $\mu\equiv\mu_B/2, \nu\equiv\mu_I/2,\nu_5\equiv\mu_{I5}/2$)
\begin{eqnarray}
\widetilde{\cal M}=\mu-\nu\tau_3-\nu_{5}\gamma^5\tau_3. \label{IV.23}
\end{eqnarray}
Obviously, the quantity which is in the square brackets of Eq. (\ref{IV.22}) is proportional to the unit operator in the $N_c$-color space.
(Below, in all numerical calculations we put $N_c=2$.) Hence,
\begin{eqnarray}
&&\det(Z)={\rm det}^{N_c}{\cal D}\equiv{\rm det}^{N_c}\left
(\begin{array}{cc}
D_{11}~, & D_{12}\\
D_{21}~, & D_{22}
\end{array}\right ),\label{IV.24}
\end{eqnarray}
where ${\cal D}$ is the $2\times 2$ matrix in the 2-dimensional flavor space (its matrix elements $D_{kl}$ are the nontrivial operators in
the 4-dimensional spinor and in the (3+1)-dimensional coordinate spaces). Using this expression for $\det(Z)$ in Eq. (\ref{IV.18}) when
$\sigma,\vec\pi,\Delta,\Delta^*$ do not depend on coordinates $x$, and taking into account the general relation (\ref{B7}) of
Appendix \ref{ApB}, we find
\begin{eqnarray}
{\cal S}_{\rm
{eff}}(\sigma,\pi_a,\Delta,\Delta^{*})~\bigg
|_{~\sigma,\vec\pi,\Delta,\Delta^{*}=\rm {const}}
&=&-\frac{\sigma^2+\vec\pi^2+|\Delta|^2}{4H}\int d^4x-\frac {iN_c}2\ln\mbox {det}{\cal D}\nonumber\\
=-\frac{\sigma^2+\vec\pi^2+|\Delta|^2}{4H}\int d^4x&-&\frac {iN_c}2\int\frac{d^4p}{(2\pi)^4}\ln\det\overline{\cal D}(p)\int d^4x,\label{IV.32}
\end{eqnarray}
where the 2$\times$2 matrix $\overline{\cal D}(p)$ is the momentum space representation of the matrix ${\cal D}$ of Eq. (\ref{IV.24}).
Its matrix elements $\overline{\cal D}_{kl}(p)$ have the following form
\begin{eqnarray}
\overline{D}_{11}(p)&=&|\Delta |^2-p^2+\mu\big[\hat p \gamma^0-\gamma^0\hat p \big ]+
(\nu+\nu_5\gamma^5)\big[\hat p \gamma^0+\gamma^0\hat p\big ]+\vec\pi^2+M^2\nonumber\\
&+&2M\gamma^0(\mu+\nu_5\gamma^5)+\mu^2-(\nu+\nu_5\gamma^5)^2
+2i\mu\gamma^0\gamma^5\pi_3+2i\nu_5\gamma^0\pi_3,\nonumber\\
\overline{D}_{22}(p)&=&|\Delta |^2-p^2+\mu\big[\hat p \gamma^0-\gamma^0\hat p \big ]
-(\nu+\nu_5\gamma^5)\big[\hat p\gamma^0+\gamma^0\hat p\big ]+\vec\pi^2+M^2\nonumber\\
&+&2M\gamma^0(\mu-\nu_5\gamma^5)+\mu^2-(\nu+\nu_5\gamma^5)^2
-2i\mu\gamma^0\gamma^5\pi_3+2i\nu_5\gamma^0\pi_3,\nonumber\\
\overline{D}_{12}(p)&=&2i\mu\gamma^0\gamma^5\big(\pi_1+i\pi_2\big)+2\nu\gamma^0\gamma^5\big(\pi_2-i\pi_1\big)=
2\gamma^0\gamma^5(\nu-\mu)[\pi_2-i\pi_1],\nonumber\\
\overline{D}_{21}(p)&=&2i\mu\gamma^0\gamma^5\big(\pi_1-i\pi_2\big)+2\nu\gamma^0\gamma^5\big(i\pi_1+\pi_2\big)=
2\gamma^0\gamma^5(\nu+\mu)[\pi_2+i\pi_1],
\label{IV.33}
\end{eqnarray}
where $M\equiv m_0+\sigma$. Using in Eq. (\ref{IV.32}) again the general relation (\ref{IV.21}), we have
\begin{eqnarray}
&&\det\overline{D}(p)\equiv {\rm det}\left
(\begin{array}{cc}
\overline{D}_{11}(p)~, & \overline{D}_{12}(p)\\
\overline{D}_{21}(p)~, & \overline{D}_{22}(p)
\end{array}\right )\nonumber\\
&=&\det\Big[-\overline{D}_{21}(p)\overline{D}_{12}(p)+\overline{D}_{21}(p)\overline{D}_{11}(p)\big(\overline{D}_{21}(p)\big)^{-1}
\overline{D}_{22}(p)\Big]\equiv\det L(p).\label{IV.34}
\end{eqnarray}
Notice that the matrix $L(p)$, i.e. the expression in square brackets of Eq. (\ref{IV.34}), is indeed a 4$\times$4 matrix in
4-dimensional spinor space only, which is composed of 4$\times$4 matrices $\overline{D}_{ij}(p)$ (see in Eq. (\ref{IV.33})).
Now, taking into account the definition (\ref{IV.19}) and using the Eqs. (\ref{IV.32})-(\ref{IV.34}), it is possible to obtain in
the mean-field approximation and at  $N_c=2$ the following expression for the TDP of the model,
\begin{eqnarray}
\Omega(M,\vec\pi,\Delta,\Delta^{*})
&=&\frac{(M-m_0)^2+\vec\pi^2+|\Delta|^2}{4H}+i\int\frac{d^4p}{(2\pi)^4}\ln\det L(p).\label{IV.35}
\end{eqnarray}
Since $\det L(p)=\lambda_1(p)\lambda_2(p)\lambda_3(p)\lambda_4(p)$, where $\lambda_i(p)$ ($i=1,...,4$) are four eigenvalues
of the 4$\times$4 matrix $L(p)$, in the following, in order to find the TDP of the model in various cases, first of all, we will find the eigenvalues of the matrix $L(p)$. Then, after integration in Eq. (\ref{IV.35}) over $p_0$, this TDP is used in some numerical
calculations with sharp three-momentum cutoff $\Lambda=657$ MeV (i.e. it is assumed below that the integration over three-momentum $\vec p$
occurs over the region $|\vec p|<\Lambda$) at $H=7.23$ GeV$^{-2}$ and $m_0=5.4$ MeV \cite{brauner1,andersen2}. In order to simplify numerical calculations, we study the phase structure of the model in the chiral limit, $m_0=0$, at the same values of $\Lambda$ and $H$.

\section{Calculation of the TDP (\ref{IV.35}) and its duality properties }

\subsection{The case $\mu\ne 0$, but $\nu=0$ and $\nu_{5}=0$}

First of all let us consider the TDP (\ref{IV.35}) in the most simple case when only $\mu\ne 0$. In this case, using any package of
analytical calculations, it is easy to establish that the 4$\times$4 matrix $L(p)$ of Eqs. (\ref{IV.34}) and (\ref{IV.35}) has a unique
fourfold degenerated eigenvalue $\lambda(p)$,
\begin{eqnarray}
\lambda(p)=\left [p_0^2-|\Delta|^2-\left (\sqrt{\vec p^2+\vec\pi^2+M^2}+\mu\right )^2\right ]\left [
p_0^2-|\Delta|^2-\left (\sqrt{\vec p^2+\vec\pi^2+M^2}-\mu\right )^2\right].\label{IV.36}
\end{eqnarray}
(It means that $L(p)$ is proportional to the identity 4$\times$4 matrix.) As a result, in this case the TDP
(\ref{IV.35}) looks like
\begin{eqnarray}
\Omega(M,\vec\pi,\Delta,\Delta^{*})
&=&\frac{(M-m_0)^2+\vec\pi^2+
|\Delta|^2}{4H}+4i\int\frac{d^4p}{(2\pi)^4}\ln\left [(p_0^2-E^2_+)(p_0^2-E_-^2)\right ],\label{IV.37}
\end{eqnarray}
where
\begin{eqnarray}
E_\pm=\sqrt{|\Delta|^2+\left (\sqrt{\vec p^2+\vec\pi^2+M^2}\pm\mu\right )^2}.
\label{IV.38}
\end{eqnarray}
Taking into account the general formula
\begin{eqnarray}
\int_{-\infty}^\infty dp_0\ln\big
((p_0+a)^2-b^2)=\mathrm{i}\pi\big(|a-b|+|a+b|\big)\label{IV.39}
\end{eqnarray}
(being true up to an infinite term independent of the real quantities $a$ and $b$),
it is possible to integrate in (\ref{IV.37}) over $p_0$. Then, the TDP takes the form
\begin{eqnarray}
\Omega(M,\vec\pi,\Delta,\Delta^{*})
&=&\frac{(M-m_0)^2+\vec\pi^2+
|\Delta|^2}{4H}-4\int\frac{d^3p}{(2\pi)^3}\big (E_++E_-\big ).
\label{IV.40}
\end{eqnarray}
(In particular, it is clear from this expression that at $m_0=0$ and $\mu=0$ the TDP (\ref{IV.40}) is a function of the quantity
$\sqrt{M^2+\vec\pi^2+|\Delta|^2}$.) We have from (\ref{IV.40}) the following gap equations ($k=1,2$; $a=1,2,3$)
\begin{eqnarray}
\frac{\partial}{\partial\Delta_k}\Omega&\equiv&
\frac{\Delta_k}{2H}-4\Delta_k\int\frac{d^3p}{(2\pi)^3}\big (\frac{1}{E_+}+\frac{1}{E_-}\big )=0,
\label{IV.401}\\
\frac{\partial}{\partial M}\Omega&\equiv&
\frac{M-m_0}{2H}-4M\int\frac{d^3p}{(2\pi)^3}\big (\frac{\sqrt{\vec p^2+\vec\pi^2+M^2}+\mu}{E_+\sqrt{\vec p^2+\vec\pi^2+M^2}}+
\frac{\sqrt{\vec p^2+\vec\pi^2+M^2}-\mu}{E_-\sqrt{\vec p^2+\vec\pi^2+M^2}}\big )=0,
\label{IV.403}\\\frac{\partial}{\partial\pi_a}\Omega&\equiv&
\frac{\pi_a}{2H}-4\pi_a\int\frac{d^3p}{(2\pi)^3}\big (\frac{\sqrt{\vec p^2+\vec\pi^2+M^2}+\mu}{E_+\sqrt{\vec p^2+\vec\pi^2+M^2}}+
\frac{\sqrt{\vec p^2+\vec\pi^2+M^2}-\mu}{E_-\sqrt{\vec p^2+\vec\pi^2+M^2}}\big )=0,
\label{IV.404}
\end{eqnarray}
where $\Delta=\Delta_1+i\Delta_2$ and $\Delta^{*}=\Delta_1-i\Delta_2$. Compairing the gap equations (\ref{IV.403}) and (\ref{IV.404}) one can see that at $m_0\ne 0$ in the global minimum point (GMP)
of the TDP (\ref{IV.40}) we always have $\pi_a=0$ ($a=1,2,3$). (Otherwise, the gap equation (\ref{IV.403}) is equivalent to the relation
$M-m_0=M$, which means that $m_0=0$. And we have a contradiction with the initial assumption that the bare quark mass $m_0$
is not zero.) Hence, at $m_0\ne 0$ and in the case when only $\mu\ne0$ (other chemical potentials are zero) we may put
$\vec\pi_a=0$. As a result we see that at $m_0\ne 0$ and $\mu\ne 0$ the phase structure of the model (\ref{IV.100}) is really described by
the TDP $\Omega(M,\pi_a=0,\Delta,\Delta^{*})$ which is a function of only three condensates,
$M$, $\Delta$ and $\Delta^*$. This conclusion is in agreement with the consideration of the phase structure of this
model at nonzero temperature and $\mu\ne 0$ in Ref. \cite{weise}, where it was assumed that pions are not condensed. More general, this statement is always true if
the eigenvalues $\lambda_i(p)$ of the matrix $L(p)$ in Eq. (\ref{IV.34}) depend on $M$ and $\vec\pi$ in the form of the variable
$(M^2+\vec\pi^2)$, as it occurs in Eq. (\ref{IV.36}).

{\bf The case $m_0\ne 0$.} In this case, as it was shown in Ref. \cite{weise}, at $\mu<\mu_c\equiv m_{\pi}/2$ the phase with
nonzero chiral condensate $\vev{\sigma}$ is arranged. Then at $\mu>\mu_c$ the spontaneous breaking of the baryon $U(1)_B$ symmetry occurs and
baryon superfluid (or diquark condensation) phase in which $\vev{|\Delta|}\ne 0$ appears.

{\bf The case $m_0=0$.} Below, just in the chiral limit, we will mainly study the properties of the NJL model (\ref{IV.100}). Therefore, notice that
at $m_0=0$ and for all values of $\mu$ such that $0<\mu<\tilde\mu_c\approx 0.88$ GeV the BSF phase is realized. Based on this phase portrait and using the dual properties of the model,
we are able to predict the phase structure of the model for other special cases of chemical potentials (see the next sections). \label{III.A}

\subsection{The case with $\mu\ne 0$, $\nu\ne 0$ and $\nu_5\ne 0$}
\label{III.B}

In the previous sections we have considered the TDP (\ref{IV.35}) of the initial model as a function of six variables (condensates),
$M$, $\vec\pi$, $\Delta$ and $\Delta^{*}$. But due to a symmetry of the model, the number of condensates may be reduced. Indeed,
at $m_0=0$ and zero chemical potentials the Lagrangian (\ref{IV.1}) is invariant under $SU(4)\times U(1)_B\times SU(2)_c$ group.
If chemical potentials are taken into consideration, then in the chiral limit the symmetry of the Lagrangian (\ref{IV.100}) reduces to
$U(1)_B\times U(1)_{I_3}\times U(1)_{AI_3}$ symmetry (plus color $SU(2)_c$, which in our consideration is not violated at all). As a result,
we see that at $m_0=0$ and nonzero chemical potentials the TDP of the model depends only on the $|\Delta|^2$,
$\pi_1^2+\pi_2^2$ and $M^2+\pi_0^2$ field combinations, correspondingly. It is important to note that the matrix $L(p)$ as well as
the one-loop contribution to the TDP, i.e. the last term in Eq. (\ref{IV.35}), do not depend on the bare mass $m_0$. Hence in any case,
both at $m_0=0$ and $m_0\ne 0$, these quantities depend only on the above mentioned combinations of bosonic fields. As a result, we
see that in the
chiral limit one can put $\pi_0=0$ and $\pi_2=0$ without loss of generality of consideration. However, at $m_0\ne 0$ and at nonzero chemical
potentials the symmetry of
the Lagrangian (\ref{IV.100}) reduces to $U_B(1)\times U_{I_3}(1)$, i.e. in this case the TDP depends on $|\Delta|^2$, $M,\pi_0$ and
$\pi_1^2+\pi_2^2$. So in this case without loss of generality we can also use the symmetry properties of the model and put $\pi_2=0$.
Moreover, since at $m_0\ne 0$ the discussion presented just after the Eq. (\ref{IV.404}) is valid, we can attract these dynamical reasons
and put $\pi_0=0$ in addition. Hence, below throughout the paper we suppose that the TDP (\ref{IV.35}) is a function of only $M,\pi_1$ and
$|\Delta|$.

In the case when $\mu\ne 0$, $\nu\ne 0$ and $\nu_5\ne 0$ the matrix $L(p)$ of Eqs. (\ref{IV.34}) and (\ref{IV.35}) has four different
eigenvalues $\lambda_{i}(p)$ (they can be found
with the help of any package of analytical calculations),
\begin{eqnarray}
&&\lambda_{1,2}(p)=N_1\pm 4\sqrt{K_1},~~\lambda_{3,4}(p)=N_2\pm 4\sqrt{K_2},
\label{IV.56}
\end{eqnarray}
where
\begin{eqnarray}
\hspace{-1cm}N_2=N_1+16\mu\nu\nu_5|\vec p|,~~K_2=K_1+8\mu\nu\nu_5|\vec p|p_0^4-8\mu\nu\nu_5|\vec p|p_0^2\big (M^2+\pi_1^2+|\Delta|^2
+|\vec p|^2+\mu^2+\nu^2-\nu_5^2\big ),&&
\label{IV.57}
\end{eqnarray}
\begin{eqnarray}
\hspace{-1cm} K_1=\nu_5^2p_0^6-2\nu_5p_0^4\Big [\nu_5 (|\Delta|^2+\pi_1^2+M^2+|\vec p|^2+\nu^2+\mu^2-\nu_5^2)+2\mu\nu|\vec p|\Big ]+
p_0^2 \Big\{
\nu_5^6-4|\vec p|\mu\nu\nu_5^3&&~~~~~~~~~~~~~~\nonumber\\
+2\nu_5^4 \big (M^2-|\Delta|^2-\pi_1^2-\nu^2-\mu^2-|\vec p|^2\big )+4\mu^2\nu^2(M^2+
|\vec p|^2)+4|\vec p|\mu\nu(|\Delta|^2+\pi_1^2+M^2+|\vec p|^2+\nu^2+\mu^2)\nu_5&&\nonumber\\
+\nu_5^2\Big [(|\Delta|^2+\pi_1^2+|\vec p|^2+\nu^2+\mu^2)^2+2|\vec p|^2M^2+M^4+2M^2(|\Delta|^2-\nu^2+\pi_1^2-\mu^2)\Big ]\Big\},&&
\label{IV.58}
\end{eqnarray}
\begin{eqnarray}
N_1&=&p_0^4-2p_0^2\Big [|\Delta|^2+\pi_1^2+M^2+|\vec p|^2+\nu^2+\mu^2-3\nu_5^2\Big ]+
\nu_5^4-2\nu_5^2\Big [|\Delta|^2+\pi_1^2+|\vec p|^2+\nu^2+\mu^2-M^2\Big ]\nonumber\\
&-&8\mu\nu\nu_5|\vec p|+\left (|\vec p|^2+M^2+\pi_1^2+|\Delta|^2-\mu^2-\nu^2\right )^2-
4\left (\mu^2\nu^2-\pi_1^2\nu^2-|\Delta|^2\mu^2\right ),\label{IV.59}
\end{eqnarray}
and the TDP (\ref{IV.35}) has the form
\begin{eqnarray}
\hspace{-1.5cm} \Omega(M,\pi_1,|\Delta|)
&=&\frac{(M-m_0)^2+\pi_1^2+|\Delta|^2}{4H}+i\int\frac{d^4p}{(2\pi)^4}\Big\{\ln\Big(\lambda_1(p)\lambda_2(p)
\Big)+\ln\Big(\lambda_3(p)\lambda_4(p)\Big )\Big\}.\label{IV.73}
\end{eqnarray}
It is easy to see from Eqs. (\ref{IV.57})-(\ref{IV.59}) that each of the eigenvalues $\lambda_i$ (\ref{IV.56}) is invariant with respect
to the so-called dual transformation ${\cal D}_1$,
\begin{eqnarray}
{\cal D}_1: ~~~~\mu\longleftrightarrow\nu,~~~\pi_1\longleftrightarrow |\Delta|.\label{IV.55}
\end{eqnarray}
As a result, the whole TDP (\ref{IV.73}) is also invariant under the transformation (\ref{IV.55}). (For the first time, the discrete symmetry
(\ref{IV.55}) of the thermodynamic potential of the NJL model with two-color quarks was noted, but only at $\nu_5=0$, in Refs. \cite{son2,andersen2}.
It was also shown there that due to this symmetry, the charged PC and BSF phases are located symmetrically on the $(\mu,\nu)$-phase diagram
of the model.)
Moreover, using any program of analytical calculations, it is possible to establish that
the quantities $\lambda_1(p)\lambda_2(p)$ and $\lambda_3(p)\lambda_4(p)$ are invariant in addition with respect to the dual discrete
transformations ${\cal D}_2$ and ${\cal D}_3$, where
\begin{eqnarray}
{\cal D}_2: ~~\mu\longleftrightarrow\nu_5,~~M\longleftrightarrow |\Delta|;~~~~~~{\cal D}_3:
~~\nu\longleftrightarrow\nu_5,~~M\longleftrightarrow \pi_1.\label{IV.60}
\end{eqnarray}
Indeed, it is possible to show that
\begin{eqnarray}
\lambda_1(p)\lambda_2(p)=p_0^8+\alpha p_0^6+\beta p_0^4+\gamma p_0^2+\delta,\label{IV.600}
\end{eqnarray}
where
\begin{eqnarray}
\alpha&=&-4\big(|\Delta|^2+\pi_1^2+M^2+|\vec p|^2+\nu^2+\mu^2+\nu_5^2\big),\nonumber\\
\beta&=&6\big(\nu^4+\mu^4+\nu_5^4\big)+6\big(|\Delta|^2+\pi_1^2+M^2+|\vec p|^2\big)^2+48\mu\nu\nu_5|\vec p|\nonumber\\
&+&4\big(|\Delta|^2+\pi_1^2+M^2+|\vec p|^2\big)\big(\nu^2+\mu^2+\nu_5^2\big)+4\big(\mu^2\nu^2+\mu^2\nu_5^2+\nu^2\nu_5^2\big)\nonumber\\
&+&8\big(\nu^2\pi_1^2+\mu^2|\Delta|^2+\nu_5^2M^2\big).\label{IV.601}
\end{eqnarray}
It can be seen from relations (\ref{IV.601}) that the coefficients $\alpha$ and $\beta$ are invariant under the dual transformations
${\cal D}_1$, ${\cal D}_2$ and ${\cal D}_3$. The exact expressions for the coefficients $\gamma$ and $\delta$ of the polynomial
(\ref{IV.600}) are too extensive, therefore they are not presented in formula (\ref{IV.601}). However, they are also invariant under the
dual transformations (\ref{IV.55}) and (\ref{IV.60}). In a similar way it is possible to show that the product $\lambda_3(p)\lambda_4(p)$ is also
invariant with respect to the ${\cal D}_1$, ${\cal D}_2$ and ${\cal D}_3$ transformations.
Hence, the quantity $\det L(p)=\lambda_1(p)\lambda_2(p)\lambda_3(p)\lambda_4(p)$ is invariant
under the dual transformations (\ref{IV.55}) and (\ref{IV.60}). (Note that separately, each eigenvalue $\lambda_i$ (\ref{IV.56}) is not
invariant under ${\cal D}_2$ and ${\cal D}_3$, only the products $\lambda_1(p)\lambda_2(p)$ and $\lambda_3(p)\lambda_4(p)$ are
invariant.) As a result we see that only in the chiral limit, i.e. when $m_0=0$, the TDP (\ref{IV.73}) is invariant under three dual
transformations ${\cal D}_1$, ${\cal D}_2$ and ${\cal D}_3$. At the physical point, i.e. when $m_0\ne 0$, it is symmetrical only under
${\cal D}_1$. Moreover, as it is shown in Appendix \ref{ApA}, in this case the invariance of the TDP under ${\cal D}_1$ transformation
is a consequence of the symmetry of the Lagrangian (\ref{IV.100}) with respect to the dual transformation (\ref{A8}).

Note that in the general case when $m_0\ne 0$ the investigation of the TDP (\ref{IV.73}) is a rather hard task. So in the following we study the NJL model only in the chiral limit, $m_0=0$, as a rule. It is possible because in general the chiral limit often is a good approximation. In particular, this conclusion is supported by the results of Ref. \cite{kkz2}, where it was shown that for the three-color
NJL model with $\mu,\nu,\nu_5$ chemical potentials and $m_0\ne 0$, the phase structure obtained in the chiral limit gives a rather good approximation for the phase portrait, except the region of low chemical potentials, smaller than the value of the half of the pion mass. Moreover, the exact dual symmetry ${\cal D}_{3}$ (\ref{IV.60}) of this massless three-color NJL model is a very good approximation at $m_0\ne 0$. Therefore,
we think that the two-color NJL model (\ref{IV.100}) can be investigated with a sufficient degree of reliability in the chiral limit, as well.
%


Let us argue that in this case it is possible to assume that for sufficiently low values of the chemical
potentials (say at $\mu,\nu,\nu_5 <1$ GeV) at the global minimum point (GMP) $(M,\pi_1,|\Delta|)$ of the TDP (\ref{IV.73}), there can be no more than one nonzero coordinates, i.e. condensates or order parameters. The argument could be that in previous investigations there have not been found
(mixed) phases with several nonzero condensates, e.g., in the three-color NJL model \cite{kkz2} and in the two-color one \cite{son2,andersen2}. But it is not overwhelmingly convincing argument anyway. There is a more persuasive argument indicating that it is the case. Dualities stand for that.
It will be discussed below that the phase structure in terms of chiral symmetry breaking and charged pion condensation is the same for two-color and three-color NJL models. The phase structure of the three-color NJL model has been studied in Refs. \cite{kkz2,kkz,kkz1+1}. And it has been shown that there is no mixed phase with simultaneously non-zero chiral and charged pion condensates. So,
one can conclude that there is no such a mixed phase in two-color case as well.  Then one can apply the duality ${\cal D}_{1}$ (\ref{IV.55}) and argue that there is no mixed phase, where there are simultaneously non-zero chiral and diquark condensates. Or use the duality ${\cal D}_{2}$ (\ref{IV.60}) and show that there is no mixed phase, where there are simultaneously non-zero pion and diquark condensates. \footnote{Strictly speaking, the statement that it is impossible to have two or three condensates that are not equal to zero in the ground state of the system refers to the points of a phase diagram that do not lie on the boundary between phases. In the last case there might occurs another situation. For example, in Ref. \cite{adhikari} it was shown in the framework of the chiral perturbation theory for two-color QCD that at $\mu_B=\mu_I$ (in Fig. 1 (a) it is the boundary between charged PC and BSF phases) there is a multi-component superfluid phase in  which two condensates are nonzero, $\pi_1\ne 0$ and $|\Delta|\ne 0$. The same is valid in the framework of the NJL model (2) as well. Indeed, it is easy to check that at $\mu_B=\mu_I$ the eigenvalues $\lambda_i$ (\ref{IV.56}) and the TDP (\ref{IV.73}) depend on the combination $\pi_1^2+|\Delta|^2$ but not on $\pi_1$ and $|\Delta|$ separately. Hence, in this case in the GMP of the TDP (\ref{IV.73}) the relations $\pi_1\ne 0$ and $|\Delta|\ne 0$ can be realized. But in the present paper we do not study the properties of the system on the boundaries between phases, so the corresponding exceptional sets of chemical potentials, e.g., such as $\mu_B=\mu_I$, are outside the scope of our consideration.}
Let us note that this does not exclude the case where all three (chiral, pion and diquark) condensates are simultaneously non-zero, but let us be honest it is quite unlikely, especially considering the fact that mixed phases are not common. So it was rigorously proved that there is no mixed phases with only two non-zero condensates and one can undisturbed assume that there are no mixed phases (mixed phase with three condensates are extremely unlikely and even if it is present it would probably
be small region in the phase diagram). To show the absence of mixed phases with two non-zero condensates without using the dualities, would require a way more time-consuming numerical calculations.

Therefore, with such a restriction on chemical potentials, in the chiral limit only four different phases can be realized in the system. (i) If GMP has the form $(M\ne 0,\pi_1=0,|\Delta|=0)$, then the chiral symmetry breaking (CSB) phase appears in the model. (ii) If it has the form $(M=0,\pi_1\ne 0,|\Delta|=0)$, the charged pion condensation (PC) phase is realized. (iii) When the GMP looks like $(M=0,\pi_1=0,|\Delta|\ne 0)$, it corresponds to the baryon superfluid (BSF) or diquark condensation phase. And finally, (iv) the GMP of the form $(M=0,\pi_1=0,|\Delta|=0)$ corresponds to a symmetrical phase with all zero condensates. Thanks to this structure of the global minimum point, we see that in the region of relatively low values of $\mu,\nu$, and $\nu_5 $ it is enough to study not the whole TDP (\ref{IV.73}), but only its projections on the condensate axis, $F_1(M)\equiv\Omega(M,\pi_1=0,|\Delta|=0)$, $F_2(\pi_1)\equiv\Omega(M=0,\pi_1,|\Delta|=0)$ and $F_3(|\Delta|)\equiv\Omega(M=0,\pi_1=0,|\Delta|)$, where (for details see Appendix \ref{ApC})
\begin{eqnarray}
F_1(M)=\frac{M^2}{4H}
&-&\sum_{\pm}\int\frac{d^3p}{(2\pi)^3}\Big [\big |\mu+\nu\pm\sqrt{M^2+(|\vec p|-\nu_5)^2}\big|
+\big |\mu-\nu\pm\sqrt{M^2+(|\vec p|+\nu_5)^2}\big|\nonumber\\&&
+\big |\mu+\nu\pm\sqrt{M^2+(|\vec p|+\nu_5)^2}\big|+
\big |\mu-\nu\pm\sqrt{M^2+(|\vec p|-\nu_5)^2}\big|\Big ],\label{IV.189}
\end{eqnarray}
\begin{eqnarray}
F_2(\pi_1)
=\frac{\pi_1^2}{4H}
&-&\sum_{\pm}\int\frac{d^3p}{(2\pi)^3}\Big [\big |\mu+\nu_5\pm\sqrt{\pi_1^2+(|\vec p|-\nu)^2}\big|
+\big |\mu-\nu_5\pm\sqrt{\pi_1^2+(|\vec p|+\nu)^2}\big|\nonumber\\&&
+\big |\mu+\nu_5\pm\sqrt{\pi_1^2+(|\vec p|+\nu)^2}\big|+
\big |\mu-\nu_5\pm\sqrt{\pi_1^2+(|\vec p|-\nu)^2}\big|\Big ],\label{IV.190}
\end{eqnarray}
\begin{eqnarray}
F_3(|\Delta|)
=\frac{|\Delta|^2}{4H}
&-&\sum_{\pm}\int\frac{d^3p}{(2\pi)^3}\Big [\big |\nu_5+\nu\pm\sqrt{|\Delta|^2+(|\vec p|-\mu)^2}\big|
+\big |\nu_5-\nu\pm\sqrt{|\Delta|^2+(|\vec p|+\mu)^2}\big|\nonumber\\&&
+\big |\nu_5+\nu\pm\sqrt{|\Delta|^2+(|\vec p|+\mu)^2}\big|+
\big |\nu_5-\nu\pm\sqrt{|\Delta|^2+(|\vec p|-\mu)^2}\big|\Big ].\label{IV.191}
\end{eqnarray}
It is important to note that the first two terms (the last two terms) in square brackets of Eqs. (\ref{IV.189})-(\ref{IV.191}) correspond to
the contribution of the eigenvalues $\lambda_1(p)$ and $\lambda_2(p)$ (to the contribution of the eigenvalues $\lambda_3(p)$ and
$\lambda_4(p)$) of Eq. (\ref{IV.56}) to the TDP (\ref{IV.73}).
Then, comparing the smallest values of these functions, we can determine the GMP of the initial TDP (\ref{IV.73}), and, therefore, the
phase in which the system is located in the chiral limit, $m_0=0$, at given values of chemical potentials.

\subsection{Consequences of dual symmetries of the TDP in some simple cases}

By the duality property (or symmetry, or relation, etc) of any model, we understand any discrete symmetry of its TDP with respect to
transformations as order parameters (in our case, condensates $M$, $\pi_1$ and $|\Delta|$) and free external parameters of the system
(these may be chemical potentials, coupling constants, etc). The presence of the dual symmetry of the model TDP means that in its phase
portrait there is some symmetry between phases with respect to the transformation of external parameters, which can greatly simplify the
construction of the full phase diagram of the system. (The invariance of the TDP (\ref{IV.40}) under the changing of a sign of the quark
chemical potential $\mu$ is the simplest example of the dual symmetry of the model (\ref{IV.100}). Due to this kind of duality, it is enough
to study the phase structure of the model only, e.g., at $\mu> 0$ when other chemical potentials are zero, etc.) Below, we investigate the
phase portrait of the model (\ref{IV.100}) in the mean-field approximation in the presence of three nonzero chemical potentials, $\mu$,
$\nu$ and $\nu_5$, in the chiral limit. In this case, the problem is greatly simplified due to the fact that the TDP (\ref{IV.73}) of the
model has three dual symmetries, ${\cal D}_1$ (\ref{IV.55}) and ${\cal D}_2$, ${\cal D}_3$ (\ref{IV.60}).

Indeed, let us suppose that $m_0=0$ and that at the point $(\mu=a$, $\nu=b$, $\nu_5=c)$ of the phase portrait 
the GMP of the TDP (\ref{IV.73}) lies, e.g., in the point of condensate space of the form $(M=A,\pi_1=0,|\Delta|=0)$, i.e. in this case
the CSB phase is realized in the system. Then, according to the symmetries ${\cal D}_2$ and ${\cal D}_3$ (\ref{IV.60}), the TDP will have the same meaning if we interchange the values of
chemical potentials and simultaneously appropriately transpose the values of the condensates. As a result we see that, e.g., at $\mu=c$,
$\nu=b$, $\nu_5=a$ and in the point $(M=0,\pi_1=0,|\Delta|=A)$ (it is the result of the action of the ${\cal D}_2$ dual transformation on
the TDP) as well as that at $\mu=a$, $\nu=c$, $\nu_5=b$ and in the point $(M=0,\pi_1=A,|\Delta|=0)$ (it is the ${\cal D}_3$ dual transformation
of the TDP) it has the initial meaning.
Moreover, it is evident that these new points of the condensate space are nothing but the GMPs of the TDP after its ${\cal D}_2$ and
${\cal D}_3$ transformations. \footnote{Assuming the opposite, that, e.g., at $\mu=c$, $\nu=b$ and $\nu_5=a$ the GMP of the TDP is
different from the point $(M=0,\pi_1=0,|\Delta|=A)$, we can apply to this TDP again the ${\cal D}_2$ transformation and
find that at the initial values $\mu=a$, $\nu=b$ and $\nu_5=c$ of chemical potentials the GMP of the TDP is not the point
$(M=A,\pi_1=0,|\Delta|=0)$, as it was supposed from the origin. } Consequently, at the points $(\mu=c, \nu=b, \nu_5=a)$ and
$(\mu=a, \nu=c, \nu_5=b)$ of the phase diagram of the model, which we call dually ${\cal D}_2$ and dually ${\cal D}_3$ conjugated
to the starting point $(\mu=a, \nu=b, \nu_5=c)$ of the phase portrait, there are BSF and charged PC phases that are respectively
dually ${\cal D}_2$ and dually ${\cal D}_3$ conjugated to the initial CSB phase of the model. Thus, knowing the phase of the model,
which is realized at some point of its phase portrait, we can predict which phases are arranged at the dually conjugated points of a phase
diagram.

At $m_0=0$ each duality transformation ${\cal D}_i$ ($i=1,2,3)$ (\ref{IV.55}) and (\ref{IV.60}) of the TDP can also be applied to an arbitrary phase
portrait of the model as a whole. In particular, it is clear that if we have a most general $(\mu,\nu,\nu_5)$-phase portrait, then the
action, e.g., of the ${\cal D}_3$ on the TDP can be understood as the following dual ${\cal D}_3$ transformation of the model
$(\mu,\nu,\nu_5)$-phase portrait. It is necessary to rename both the diagram axes and phases in such a way, that $\nu\leftrightarrow\nu_5$
and CSB$\leftrightarrow$charged PC. At the same time the $\mu$-axis and BSF and symmetrical phases should not change their names.
It is evident that after such ${\cal D}_3$ transformation the $(\mu,\nu,\nu_5)$-phase diagram is mapped to
itself, i.e. the most general $(\mu,\nu,\nu_5)$-phase portrait is self-${\cal D}_3$-dual. Furthermore, the self-${\cal D}_3$-duality of the
$(\mu,\nu,\nu_5)$-phase portrait means that in the three-dimensional $(\mu,\nu,\nu_5)$ space the regions of the CSB and charged PC phases
are arranged mirror-symmetrically with respect to the plane $\nu=\nu_5$ of this space.
In a similar way it is possible to describe the action
of other, ${\cal D}_1$ and ${\cal D}_2$, duality transformations on the $(\mu,\nu,\nu_5)$-phase portrait of the model, which is, of course,
invariant, or self-dual, under these mappings. But different cross-sections of the full $(\mu,\nu,\nu_5)$-phase diagram, e.g., the
$(\mu,\nu)$-phase portrait at some fixed value of $\nu_5$, are not invariant, in general, under the action of dual transformations (see
below for some examples). Finally, note that under any ${\cal D}_i$ ($i=1,2,3)$ transformation the symmetrical phase remains
intact, i.e. it does not change its position on the phase diagram.

\begin{figure}
\hspace{-1cm}\includegraphics[width=1.0\textwidth]{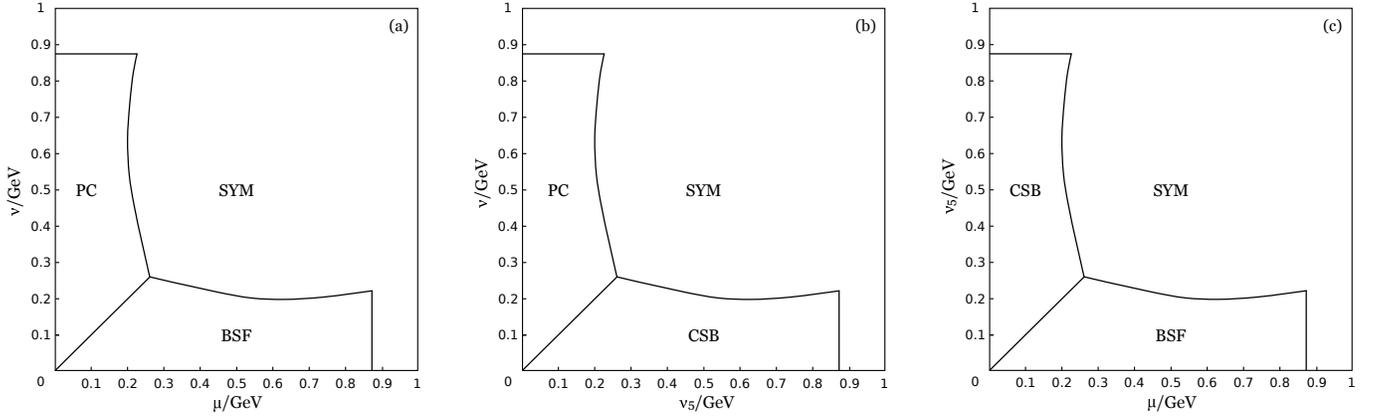}
 \caption{ (a) $(\mu,\nu)$-phase diagram at $\nu_{5}=0$. (b) $(\nu_5,\nu)$-phase diagram at $\mu=0$. It is dually ${\cal D}_2$ conjugated
 to a diagram  of Fig. 1(a). (c) $(\mu,\nu_5)$-phase diagram at $\nu=0$. It is dually ${\cal D}_3$ conjugated
 to a diagram  of Fig. 1(a). In all panels BSF means the baryon superfluid (or diquark condensation) phase in which
 $M=0$, $\pi_1= 0$, $|\Delta|\ne 0$. PC is the shorthand for the charged pion condensation phase with $M=0$, $\pi_1\ne 0$,
 $|\Delta|= 0$. In the chiral symmetry breaking (CSB) phase $M\ne 0$, $\pi_1= 0$, $|\Delta|= 0$. SYM is the
 notation for the symmetrical phase, where $M=0$, $\pi_1= 0$, $|\Delta|= 0$.}
\end{figure}
As a result, based on this mechanism of dual transformation of phase diagrams and using previously obtained particular (simple) phase portraits
of the NJL model (\ref{IV.100}) with two quark colors, it is possible to describe its phase structure at qualitatively different chemical
potential contents at $m_0=0$. Namely,
\begin{itemize}
\item 
For example, as it is noted at the end of subsection \ref{III.A}, the phase structure of the NJL model (\ref{IV.100}), when only
quark chemical potential $\mu$ is nonzero, was studied in Ref. \cite{weise} at $m_0\ne 0$ and nonzero temperature $T$. Reducing this analysis
to the case $m_0=0$, one can see that at $0<\mu<0.88$ GeV and $T=0$ the BSF phase is realized in the model. It means that at $\mu>0$, $\nu=0$ and
$\nu_5=0$ the condensates are the following, $(M=0,\pi_1=0,|\Delta|\ne 0)$. Applying to this phase diagram the ${\cal D}_1$ duality
tranformation (\ref{IV.55}), we replace $\mu$ by $\nu$ and BSF phase by charged PC phase and find the phase portrait of the model at $\nu>0$, $\mu=0$
and $\nu_5=0$. Hence, we see that if only isospin asymmetry is presented in the model (i.e. only $\nu\ne 0$), then in the chiral
limit and for sufficiently low values of $\nu$ the charged PC phase is observed. However, acting on the original phase portrait by duality transformation
${\cal D}_2$ (\ref{IV.60}), one can find that if $\nu_5>0$ (and small enough), then usual CSB phase is realized in the model in the chiral limit.
\item 
There is another simple phase diagram of the model (\ref{IV.100}) from which it is possible to find its phase structure
for qualitatively different chemical potential sets, using only dual transformations of this phase diagram (and without any numerical calculations).
Earlier, the $(\mu,\nu)$-phase portrait of the model was obtained in Refs. \cite{son2,andersen2} at $\nu_5=0$ and $m_0\ne0$. In Fig. 1(a) this phase
portrait is presented in the chiral limit, and it is clear that there, due to the ${\cal D}_1$-invariance (\ref{IV.55}) of the model TDP (\ref{IV.73}), the
BSF and charged PC phases of the model are arranged mirror-symmetrically with respect to the line $\mu=\nu$ (it is one of the conclusions of
the papers \cite{son2,andersen2}), i.e. this diagram is a self-dual with respect to ${\cal D}_1$ mapping.  Then, performing in each point
of this diagram the ${\cal D}_2$ dual transformation of the TDP, one can get the so-called ${\cal D}_2$ dual conjugation (or mapping) of
this $(\mu,\nu)$-phase portrait at $\nu_5=0$, which is no more than the $(\nu_5,\nu)$-phase portrait of the model at $\mu=0$ (see in Fig. 1(b)).
In a similar way, applying to the diagram of Fig. 1(a) the ${\cal D}_3$ dual transformation, one can obtain its ${\cal D}_3$ dual conjugation,
i.e.  the $(\mu,\nu_5)$-phase portrait of the model at $\nu=0$, where instead of the charged PC the CSB phase appears. And
mirror-symmetrically to it (with respect to the line $\nu_5=\mu$) the BSF phase is arranged. Note that under any dual
transformation ${\cal D}_i$ the position of the symmetrical phase is not changed.
\end{itemize}
Taking into account these simple phase diagrams, e.g., presented in Fig. 1, it is possible to say that different phenomena are generated
by different chemical potentials. So, charged PC is caused by isospin chemical potential $\mu_I\equiv 2\nu$, baryon chemical potential
$\mu_B\equiv 2\mu$ leads to the diquark condensation in the system, whereas at non-zero chiral isospin chemical potential
$\mu_{I5}\equiv 2\nu_5$ the chiral symmetry gets broken. \label{III.C}

\section{Phase structure in the general case $\mu\ne 0, \nu\ne 0$ and $\nu_5\ne 0$}

\begin{figure}
\hspace{-1cm}\includegraphics[width=1.0\textwidth]{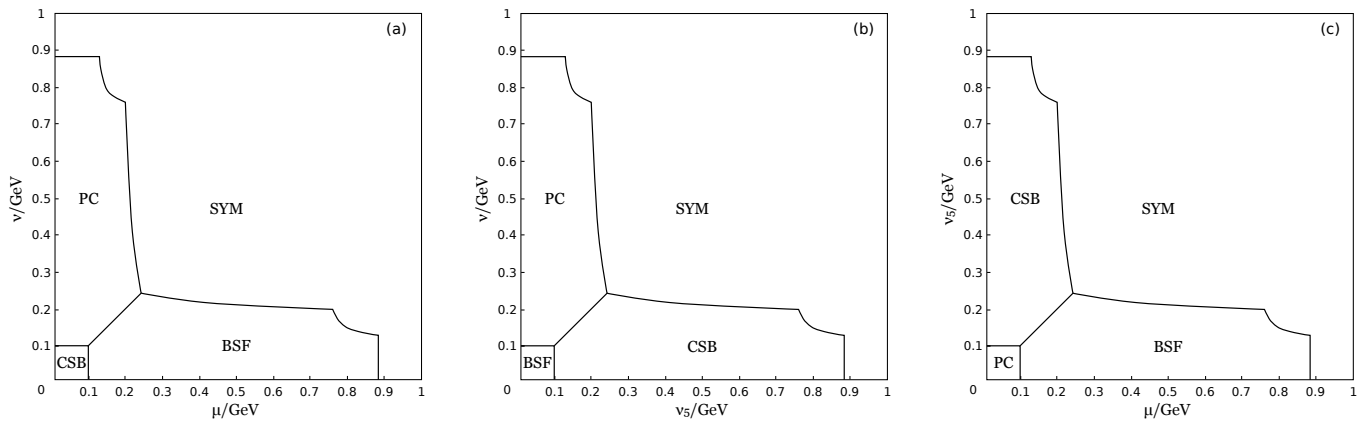}
 \caption{ (a) $(\mu,\nu)$-phase diagram at $\nu_{5}=0.1$ GeV. (b) $(\nu_5,\nu)$-phase diagram at $\mu=0.1$ GeV. (c) $(\mu,\nu_5)$-phase
 diagram at $\nu=0.1$ GeV. All the notations are the same as in previous Fig. 1.
}
\end{figure}

In the present section we study the general $(\mu,\nu,\nu_5)$-phase portrait of the model in the chiral limit. In this case the investigation is
simplified due to the dual symmetries ${\cal D}_i$ (\ref{IV.55}) and (\ref{IV.60}) of the TDP (\ref{IV.73}). Moreover, as it was noted in
subsection \ref{III.B}, to find the GMP of the TDP at $m_0=0$ and sufficiently small values of chemical potentials, $\mu,\nu,\nu_5
\lesssim 1$ GeV, it is enough to compare the least values of the projections (\ref{IV.189})-(\ref{IV.191}) of the TDP (\ref{IV.73}) on the
condensate axis. Just the behaviour of the GMP vs chemical potentials determines the phase structure of the model. As a result we see
that in the region, where $\mu$, $\nu$ and $\nu_{5}$ are smaller than 1 GeV, only four phases listed in subsection \ref{III.B}, i.e. CSB,
charged PC, BSF and symmetrical phases, can appear in the general
$(\mu,\nu,\nu_5)$-phase diagram of the model. 
But even with this assumption, which greatly simplifies the numerical calculations, the finding of a complete phase portrait of the model
is a quite difficult task. So, in order to obtain a more deep understanding of the phase diagram as well as to get greater visibility of it,
it is very convenient to consider different cross-sections of this general $(\mu,\nu,\nu_{5})$-phase portrait by the planes of the form
$\mu= const$, $\nu= const$ and $\nu_5= const$.

Below, these different cross-sections of the most general $(\mu,\nu,\nu_{5})$-phase portrait will be presented in the chiral limit.
Moreover, the solution of the problem is greatly facilitated because of the dual symmetries of the thermodynamic potential. Indeed, it is
sufficient to construct, for example, $(\mu,\nu)$-phase diagrams for some fixed values of $\nu_5$, and then apply duality transformations
${\cal D}_{2,3}$ (\ref{IV.60}) to them according to the scheme described in the subsection \ref{III.C}. As a result, one can obtain the
missing $(\nu,\nu_5)$- and $(\mu,\nu_5)$-phase diagrams at the corresponding fixed values of $\mu$ and $\nu$, respectively.

$\bullet$ We begin the discussion of the general phase portrait of the model with several $(\mu,\nu)$-diagrams corresponding to fixed rather low
values of $\nu_5\lesssim 2.5$ GeV (see Figs. 1(a), 2(a) and 3(a,b)). Their characteristic feature 
is the fact
that the diquark condensation BSF phase is located in the region, where $\mu>\nu,\nu_5$, and, in contrast, the charged PC phase is located at
$\nu>\mu,\nu_5$. For simplicity, let us assume that $\nu_5\approx 0$. Then this property of the model can be explained qualitatively using
the same arguments as in Ref. \cite{andersen2}, where the phase structure of this model has been investigated at $\nu_5=0$. Indeed, if
$\mu>\nu$, then both $u$ and $d$ quarks form their own Fermi seas. As a result, the system gives the possibilities of the formation of $ud$
particle-particle Cooper pairs with quantum numbers of auxiliary field $\Delta (x)$ (see in Eq. (\ref{IV.3})), their
condensation, and therefore of the formation of the ground state of the BSF phase. At $\nu>\mu$, $u$ and $\bar d$ quarks form Fermi seas.
This leads both to the possibility of the formation and condensation of $u\bar d$ Cooper pairs, and to the formation of the charged PC phase in
the system.

$\bullet$  The situation is drastically changed in the case when $\nu_5$ is rather large, e.g., at $\nu_5$ greater than 0.25-0.3 GeV.
In this case the
typical $(\mu,\nu)$-phase portrait is presented in Fig 4(a), where we fix $\nu_5$ to be 0.6 GeV. But to get the idea of the phase structure
at other rather high fixed values of $\nu_5$, one can imagine that the charged PC phase (the BSF phase) of this diagram has the form of a
boot sole pointing by its tip to the value $\mu=\nu_5=0.6$ GeV of the $\mu$-axis (to the value $\nu=\nu_5=0.6$ GeV of the $\nu$-axis).
(And in the figure, these ``boot soles'' intersect with each other, as well as with the strip of the CSB phase along the line $\mu=\nu$.)
Then, to find (approximately) the $(\mu,\nu)$-phase diagram at another rather high value of $\nu_5\ne 0.6$ GeV, one should, starting from
the diagram of Fig. 4(a), simply shift the ``boot sole'' of the charged
PC phase parallel to the $\mu$-axis in a position, in which its tip is directed to the point of this axis, where $\mu=\nu_5$. In a similar
way the BSF phase should be shifted along the $\nu$-axis. Note, that already in Figs. 3(a,b) one can see the beginning of the process of
forming the charged PC and BSF phases in the form of the ``boot soles''. At rather large value of $\nu_5$, say at $\nu_5=0.4$ GeV, a small
part of the BSF phase, a tip of the ``boot sole'', penetrates into the region where $\mu<\nu$ (see in Fig. 3(c)). And at even higher values of
$\nu_5$ in all the $(\mu,\nu)$-phase diagrams obtained in this way, the BSF phase
(the charged PC phase) will be located mainly in the region $\mu<\nu\approx\nu_5$ (in the region $\nu<\mu\approx\nu_5$), what is not in
agreement with the phase diagrams of Figs. 1(a), 2(a) and 3(a,b) for fairly small $\nu_5$, where the BSF phase is located mostly at $\mu>\nu$.
\begin{figure}
\hspace{-1cm}\includegraphics[width=1.0\textwidth]{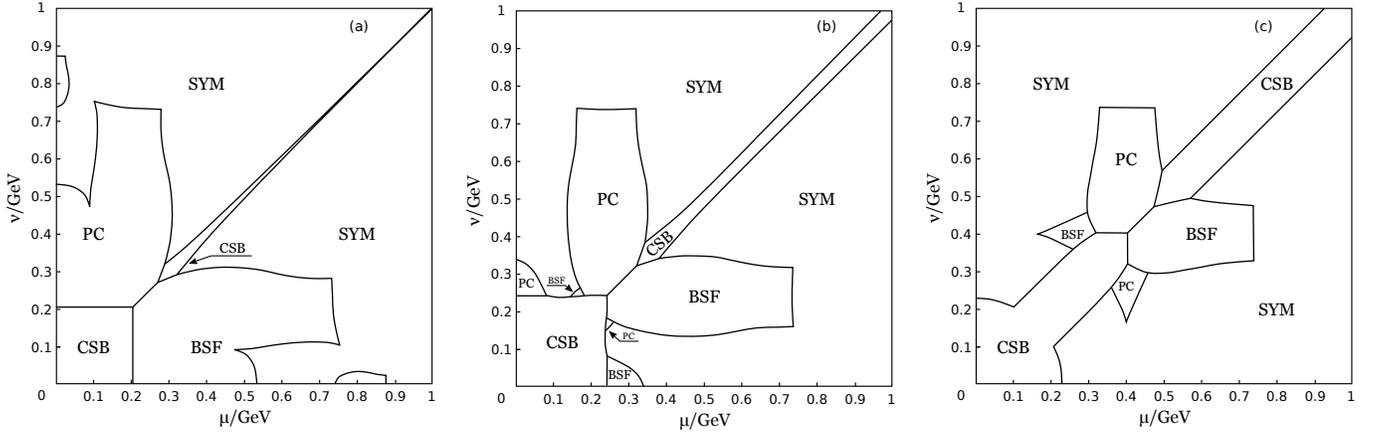}
 \caption{ (a) $(\mu,\nu)$-phase diagram at $\nu_{5}=0.2$ GeV. (b) $(\mu,\nu)$-phase diagram at $\nu_5=0.24$ GeV. (c) $(\mu,\nu)$-phase
 diagram at $\nu_5=0.4$ GeV. All the notations are the same as in Fig. 1.
}
\end{figure}

To explain this qualitative difference in the phase diagrams, it must be borne in mind that for large values of $\nu_5$, the Fermi seas and
surfaces for left- and right-handed quarks of the same flavor differ significantly from each other. Indeed, it was shown in our previous paper
\cite{kkz} that in the presence of $\mu,\nu,\nu_5$ the chemical potentials of right- and left-handed $u$ and $d$ quarks look like
\begin{eqnarray}
\mu_{uR}=\mu+\nu+\nu_5,~~\mu_{uL}=\mu+\nu-\nu_5,~~\mu_{dR}=
\mu-\nu-\nu_5,~~\mu_{dL}=\mu-\nu+\nu_5.
\label{IV.B3}
\end{eqnarray}
It is clear from (\ref{IV.B3}) that at $\mu<\nu\approx\nu_5$ the chemical potential $\mu_{dR}$ of the $d_R$ quarks is negative. So the Fermi
sea with positive energies can be formed by charge conjugated $d_R^c$ quarks, and just the $d_R^c$ quark with some momentum $\vec p$ might be
born above the corresponding Fermi surface. The chemical potential $\mu_{uR}$ of right-handed $u_R$ quarks is positive, so their Fermi sea
also exists, over the surface of which the $u_R$ quark with the same momentum $\vec p$ (as for the $d_R^c$ quark) can also be born. But
in the last case inside the $u_R$-Fermi sea the hole $\bar u_R$ with the momentum $(-\vec p)$ appears, and the $\bar u_R d_R^c$ particle-hole
Cooper pair with quantum numbers of the auxiliary field $\Delta^* (x)$ (see in Eq. (\ref{IV.3})) can be created. Condensation of these Cooper
pairs leads to a rearrangement of the ground state of the system and its transition to the BSF phase. Note once more that in the region
$\mu>\nu,\nu_5$ and rather small values of $\nu_5$ (as, e.g., in Figs. 2(a) and 3(a,b)) the diquark condensation BSF phase appears due to a
particle-particle Cooper pairing, whereas at $\mu<\nu\approx\nu_5$ and rather high values of $\nu_5$ (see in Fig. 4(a)) the BSF phase exists
due to a particle-hole Cooper pairing mechanism.

In each of Figs. 1(a), 2(a), 3 and 4(a) it is easy to see the duality ${\cal D}_1$ (\ref{IV.55}) between charged PC and BSF phases, i.e. their
symmetrical arrangement with respect to the line $\mu=\nu$. Moreover, it is clear, for example, from Fig. 4(a) that in quark matter, which is
characterized by chiral imbalance with $\nu_5=const$, the diquark condensation BSF phase can appear even at rather low values of quark
number chemical potential $\mu\gtrsim 0.2$ GeV, but only under condition that $\nu\approx \nu_5$. In contrast, in this case the charged PC
phenomenon can be realized at rather small values of $\nu$ if $\mu\approx\nu_5$. This means that in order for a BSF phase with diquark
condensate to appear in dense quark matter with chiral imbalance ($\nu_5\ne 0$), an isotopic imbalance with $\nu\approx\nu_5$ of this dense
medium is also necessary.

$\bullet$ Applying to these $(\mu,\nu)$-phase portraits (each at some fixed value of $\nu_5$) the duality transformations ${\cal D}_2$
and ${\cal D}_3$ (\ref{IV.60}), one can find the corresponding dually conjugated $(\nu,\nu_5)$- and $(\mu,\nu_5)$-phase diagrams at
corresponding fixed values of $\mu$ and $\nu$, respectively (see, e.g., in Figs. 2 and 4). So in Fig. 4 we present the $(\mu,\nu)$-phase
diagram at fixed $\nu_5=0.6$ GeV as well as its dual ${\cal D}_2$ mapping, i.e. the $(\nu,\nu_5)$-diagram at fixed $\mu=0.6$ GeV, and its
dual ${\cal D}_3$ mapping, which is the $(\mu,\nu_5)$-diagram at fixed $\nu=0.6$ GeV. Precisely the diagram of Fig. 4(b) that most clearly
supports the assertion of the previous
paragraph that just at $\nu\approx\nu_5$ the diquark condensation BSF phase can be generated in quark matter for a rather wide interval
of $\mu$. Moreover, it is clear from Fig. 4(c) that for a wide interval of $\nu$ the charged PC phase can be realized in the system if
$\mu\approx\nu_5$. And in the region, where $\mu\approx\nu\approx\nu_5$, a rather complicated interplay between BSF, CSB and charged PC
phases takes place demonstrating a very rich phase structure of the two-color NJL model. In this region several first order phase
transitions can occur (see Fig. 5, where the behavior of the condensates vs chemical potentials are presented).
It follows, e.g., from Fig. 5 (left panel) that at fixed $\nu=0.57$ GeV, $\nu_5=0.6$ GeV
and $\mu$ less than 0.49 GeV, the BSF phase is realized in the system. Then, with the growth of $\mu$ (in this case one goes deeper in
the neutron star, or baryon density increases in heavy ion collision, etc), the first order phase transition takes place to the CSB phase,
and then, at even larger $\mu$, there is a second first order phase transition to the charged PC phase, after which the system
returns to the BSF state again. In Fig. 5 (right panel), which is dually ${\cal D}_1$ conjugated to the left panel of the same figure,
the behavior of the gaps vs $\nu$ are presented. It can be be used to simulate other, dually ${\cal D}_1$ conjugate modes of physical
processes in dense baryonic medium. Moreover, the right and left panels of Fig. 5 demonstrate the fact that under the  dual ${\cal D}_1$
transformation the diquark $|\Delta|$ and charged pion $\pi_1$ condensates pass one into another.
\begin{figure}
\hspace{-1cm}\includegraphics[width=1.0\textwidth]{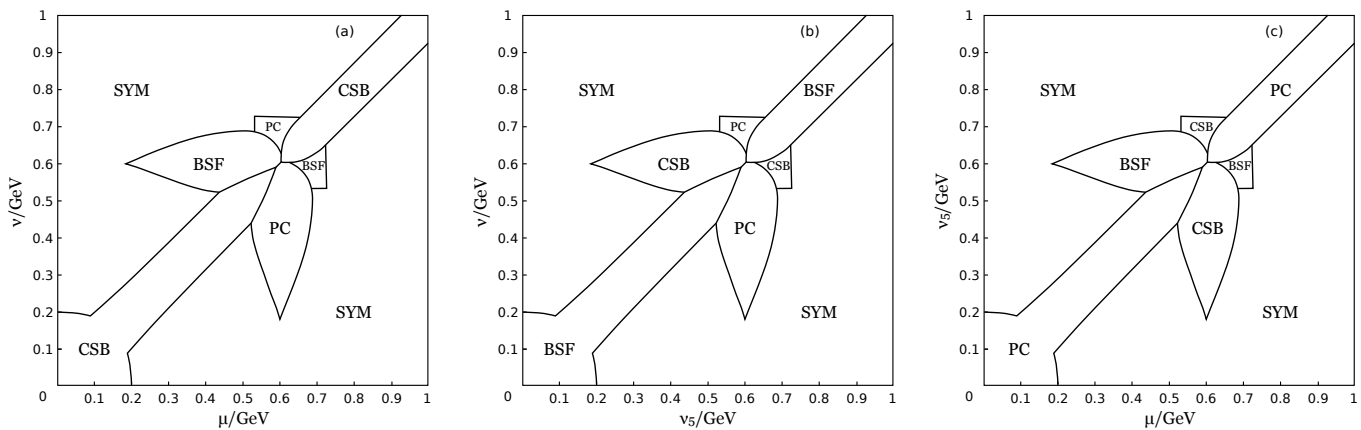}
 \caption{  (a) $(\mu,\nu)$-phase diagram at $\nu_{5}=0.6$ GeV. (b) $(\nu_5,\nu)$-phase diagram at $\mu=0.6$ GeV. (c) $(\mu,\nu_5)$-phase
 diagram at $\nu=0.6$ GeV. All the notations are the same as in Fig. 1.}
\end{figure}

$\bullet$ Finally, we would like to note another interesting feature of all phase portraits in Figs. 1-4, and the two-color NJL
model (\ref{IV.100}) itself. In Ref. \cite{kkz} the phase structure of the NJL model with three-color quarks was studied in the
mean-field approximation and the same set of chemical potentials, $\mu,\nu,\nu_5$. There, in contrast with the two-color NJL model
(\ref{IV.100}), only the $\bar q q$ interaction channel was taken into consideration with coupling constant $G$, where
$G=15.03$ GeV$^{-2}$ (i.e. the diquark $qq$ channel was ignored). Moreover, it was shown there that in this model the duality ${\cal D}_3$
(\ref{IV.60}) between CSB and charged PC phases exists (in the chiral limit, $m_0=0$). It turns out that at $G=2 H$ the TDP of the three-color
NJL model of Ref. \cite{kkz} is the half of the TDP (\ref{IV.35})-(\ref{IV.73}) of the present two-color NJL model, if the diquark condensate
$\Delta$ is zero (see the remark below Eq. (\ref{IV.184}) of Appendix \ref{ApC}), and it can be viewed as sometimes one divides coupling
constant $G$ by number of colors. (Indeed, in Ref. \cite{kkz} in the three-color considerations the coupling constant $G$ was divided by
number of colors. However, in this paper the  coupling constant $H$ was not divided by $N_{c}$ in order to be consistent and have the
same fit of parameters as in Ref. \cite{andersen2}.)
In this case, i.e. at $G=2H$, it is clear that if for some set $\mu,\nu,\nu_5$ of chemical potential values the phase
with $\Delta=0$ is observed in the framework of the two-color NJL model (\ref{IV.100}), then the same phase is realized in the three-color NJL model of Ref.
\cite{kkz} (but generally speaking, not vice versa).

Let us discuss the fit of the model parameters in more detail (see also Appendix \ref{ApC}). Note that the fit in Refs. \cite{andersen2, brauner1}, which is also used in
this paper, was constructed from the fit of three color case by the following $N_{c}$ scaling of physical quantities
$f_{\pi}\sim \sqrt{N_{c}}$ and $\langle \bar{q} q \rangle \sim N_{c}$. It can be easily shown that it is equivalent to the scaling of
coupling constant $G\sim 1/N_{c}$. 
So from the fit used in three-color case of Ref. \cite{kkz}, one could directly obtain the fit for two-color NJL model (\ref{IV.100}),
$H=G/2=7.515$ GeV$^{-2}$ and $\Lambda=650$ MeV. However, in order to be consistent with earlier investigations, we use in all numerical calculations throughout the paper
the fit employed in the previous considerations (see in Refs. \cite{andersen2, brauner1}) of the two-color case, $H=7.23$ GeV$^{-2}$ and
$\Lambda=657$ MeV (compare with the parameter fixing in the two-color NJL model under consideration given just after Eq. (\ref{IV.35})) that corresponds
to $f_{\pi}=75.4$ MeV and $-\langle \bar{q} q \rangle^{\frac{1}{3}}=218$ MeV. Nevertheless, a simple comparison shows that in this case
the relation $2H=G$ between the coupling constants of these two NJL models is performed with a rather high accuracy.
Therefore, it can be expected that in each of the diagrams of Figs. 1-4, the same phase structure is realized outside the region occupied
by the BSF phase as predicted by the three-color NJL model of Ref. \cite{kkz}. Comparing the corresponding phase portraits obtained in
these two models, we see that this is the case.

\begin{figure}
\hspace{-1cm}\includegraphics[width=1.0\textwidth]{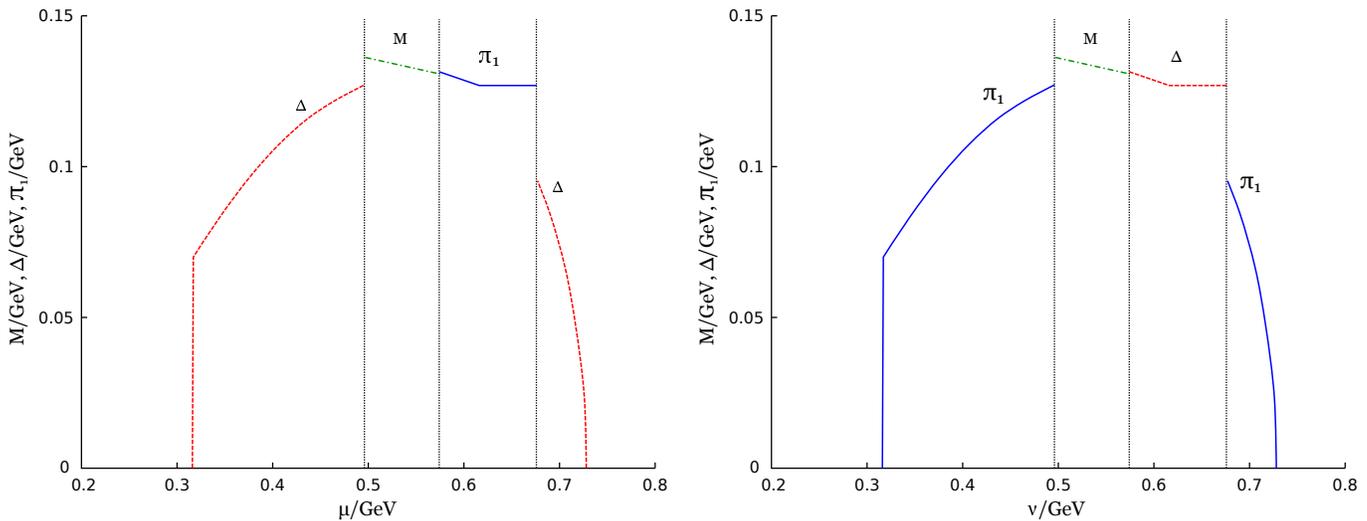}
 \caption{  Left panel: The gaps $M$ (dash-dotted line), $\pi_1$ (solid line) and $|\Delta|$ (dashed line) as a function of $\mu$ at $\nu=0.57$ GeV and $\nu_5=0.6$ GeV.
  Right panel: The gaps $M$, $\pi_1$ and $|\Delta|$ as a function of $\nu$ at $\mu=0.57$ GeV and $\nu_5=0.6$ GeV.
 Other notations are the same as in Fig. 1.}
\end{figure}

\section{Summary and Conclusions}

It is well known that at low temperatures and rather high baryon densities the appearance of the CSC (or diquark condensation) phenomenon in quark matter is expected \cite{alford}. The influence of chiral imbalance (when chiral chemical potential $\mu_5$ is nonzero) of dense cold
quark matter on CSC was investigated, e.g., in Ref. \cite{cao}, where it was shown that an increase in $\mu_5$, the CSC phase in quark matter is suppressed. This effect can be explained by the fact that the chiral chemical potential $\mu_5$ is a catalyst for spontaneous breaking of chiral symmetry \cite{braguta}. However, there is another mechanism of chiral asymmetry, which can be realized in dense quark matter in strong magnetic fields (see the discussion in Introduction) and is characterized by the chiral isospin chemical potential $\mu_{I5}$.
And it was shown that chiral isospin imbalance (when $\mu_{I5}\ne 0$) promotes the generation of the charged PC phenomenon in dense quark matter \cite{kkz2,kkz,kkz1+1} (see also the recent reviews \cite{zhokhov} on the topic). Of course, this result can be trusted in the region of relatively low baryon densities of quark matter. What happens at high densities when it is necessary to take CSC into account remains unclear.

In the present paper an interplay between various physical phenomena, CSB, charged PC and diquark condensation, in dense ($\mu_B\equiv 2\mu\ne 0$) quark matter with isospin ($\mu_I\equiv 2\nu\ne 0$) and chiral isospin ($\mu_{I5}\equiv 2\nu_5\ne 0$) imbalances is investigated at zero temperature. To simplify the consideration, we study the problem within the framework of the NJL model (\ref{IV.1})-(\ref{IV.100}), in which quarks are two color. In addition, a diquark interaction channel is also taken into account in this model and, therefore, its thermodynamic potential (TDP) $\Omega(M,\pi_1,\Delta)$ (\ref{IV.73}) depends on three order parameters (or condensates), chiral $M$, charged pion $\pi_1$, as well as diquark
$\Delta$. The last, diquark condensate, both in two-color QCD and in two-color NJL model is a colorless and electrically neutral quantity (if electric charges of quarks $q_{u,d}=\pm 1/2$, as in Ref. \cite{ramos}) with nonzero baryon charge. So if in the ground state of the system the diquark condensate $\Delta$ is nonzero, then the baryon superfluid (BSF) phase is realized (it is an analog of helium superfluidity), and the baryon $U(1)_B$ symmetry of the model is spontaneously broken. However, if $q_{u}=2/3$ and $q_{d}=-1/3$, as in Ref. \cite{andersen3}, then diquarks are colorless baryons with electric charge 1/3. Their condensation leads to a phase with spontaneous breaking of the baryon $U(1)_B$ and electromagnetic $U(1)_Q$ symmetries. This phase resembles the color superconductive phase of real QCD (in three-color QCD the diquarks have in addition color charge \cite{alford}.) Since in the present consideration the electromagnetic interaction is not taken into account, the results, i.e. the phase structure of the model (1)-(2), do not depend on which of these two ways of fixing the electric charges of quarks is chosen. So in our paper the diquark condensation phase is denoted by BSF phase, in each case.

We have calculated the TDP of this two-color NJL model in the mean-field approximation and found that in the chiral limit (bare mass of quarks is zero) it is symmetric under three discrete ${\cal D}_{1,2,3}$ (\ref{IV.55}) and (\ref{IV.60}) simultaneous transformations of both condensates and chemical potentials. Finally note that the consideration has been performed within the assumption, which is confirmed by a numerical analysis in the region of sufficiently small values of $\mu,\nu,\nu_5 <1$ GeV, that in the chiral limit there is no mixed phase. It means that in the global minimum point of the TDP two or more condensates cannot have nonzero values (see the discussion in Section \ref{III.B}).

Let us summarize the main results on our investigation of dense quark matter properties in the presence of isospin and chiral isospin imbalances within the framework of the two-color NJL model (\ref{IV.100}) with $\bar qq$ and $qq$ interaction channels (quarks are massless).

\begin{itemize}
\item Considering the $(\mu,\nu,\nu_{5})$-phase portrait of two-color quark matter as a whole, we see that for $m_0=0$ it has a surprising
(dual) symmetry between CSB, charged PC and BSF phenomena. It is a consequence of the invariance of the TDP (\ref{IV.73}) with respect to
three discrete ${\cal D}_{1,2,3}$ (\ref{IV.55}) and (\ref{IV.60}) transformations. (One of them, ${\cal D}_{3}$ is similar to the one observed
in the three-color case ($N_{c}=3$), others exist only when one has two colors.) In the simplest phase portraits, this dual symmetry
manifests itself in the symmetric arrangement of different phases in phase diagrams. For example, in Fig. 4(a) the charged PC and BSF phases
are mirror symmetric to each other with respect to the line $\mu=\nu$, which is a consequence of the dual invariance ${\cal D}_{1}$
(\ref{IV.55}) of the thermodynamic potential of the system, etc.

\item Relying on the dual symmetries ${\cal D}_{1,2,3}$ of the thermodynamic potential, one can obtain qualitatively different phase portraits
of the system without any additional numerical calculations. For example, from different $(\mu,\nu)$-phase diagrams, each of which with a
fixed value of $\nu_5$, it is possible to obtain the corresponding $(\nu,\nu_5)$- and $(\mu,\nu_5)$-phase portraits of the model which are the
dually ${\cal D}_{2}$ and dually ${\cal D}_{3}$ conjugated to the initial $(\mu,\nu)$-phase diagrams (see in Figs. 1, 2 and 4), etc.
The corresponding technical procedure for constructing a dual mapping of a phase diagram is described in section III C.

\item Based on the duality properties of the phase portrait of the model (\ref{IV.100}), we have shown that in quark matter composed
of two-color quarks, a mixed phase, i.e. phase with two non-zero simultaneously condensates, cannot be realized (see the discussion in the
section \ref{III.B}). Knowing this greatly simplifies the numeric calculations
making them several orders of magnitude less machine-time consuming.
\end{itemize}
The found dualities are very interesting properties of the phase structure and greatly simplify the investigation of the phase diagram.
The dualities interconnect the whole phase diagram (different parts of the phase diagram) and the knowledge of only a part of the phase
diagram can give us the knowledge about the whole phase structure. 
\begin{itemize}
\item It has been shown that in a sense each phenomena is inextricably linked with the corresponding chemical potential: $\mu_I$ is linked
with charged PC phenomenon and it leads to its appearance in the system, BSF (or diquark condensation) phenomenon is caused by
$\mu_B$ (in three-color case the CSC appears at large baryon chemical potential),
whereas chiral isospin chemical potential is connected with CSB phenomenon (catalysis of chiral symmetry breaking by chiral imbalance
shown as in three-color as well as in  two-color cases).
This picture can be clearly seen in the range of chemical potentials smaller than 200-300 MeV (see, e.g., the phase diagrams of Figs. 1, 2, 3(a)). At even larger chemical potentials picture
changes and becomes more rich and new interesting features appears (see below). 

\item We show that in quark matter with rather high baryon density the generation of the charged PC phase by $\nu_{5}$ is not prohibited
 at all by the processes of  diquark condensation (see in Figs. 1-4). Moreover, it can be seen from these phase diagrams that at low values
 of $\nu_{5}$ the charged PC phase is located in the region $\nu>\mu$ (see, e.g., in Figs. 1-3), whereas at large $\nu_{5}$ it can appear even
 at $\nu<\mu$ (see in Figs. 3-4).

 \item And vice versa, as it is clear from the same diagrams, at rather high values of $\nu_{5}$ the BSF phase can appear even at sufficiently
low values of $\mu$ compared with $\nu$, i.e., in a sense, chiral isospin asymmetry of quark matter with isospin density promotes the formation of diquark condensate and shifts the
BSF phase to the region of low baryon densities $\nu>\mu$, whereas at low $\nu_{5}$ the BSF phase is realized in the region where $\nu<\mu$. Chiral imbalance leads to the generation of BSF phase in quark matter with non-zero isospin density (if isospin density is larger than the baryon density). If there is no chiral imbalance one expects PC phase to appear in this regime (if $\nu>\mu$).

\item It is interesting also to note that outside of the BSF regions of all phase diagrams in Figs. 1-4, the predictions of the phase
structure, made in the present two-color NJL model (\ref{IV.100}), correspond to a phase structure which is observed in the framework of the
three-color NJL model \cite{kkz2,kkz,kkz1+1} without a diquark channel (see the remark at the end of Sec. IV). The converse is generally not
true, i.e. if in the NJL model with three-color quarks and without a diquark channel, some phase is realized (it can be CSB, charged PC or
symmetric phase) at some set $\mu,\nu,\nu_5\ne 0$, then this does not mean that in the 2-color NJL model (\ref{IV.100}) approach this phase
will also be realized at the same values of chemical potentials. The appearance of the phase with diquark condensation $\Delta\ne 0$
is also possible, which requires additional consideration.
\end{itemize}

As it was noted above, the phase diagram in the two-color NJL model at nonzero $\mu$, $\nu$ and $\nu_5$ is highly symmetric due to the existent dualities. One can also 
notice that the phase diagram investigated in the three color case \cite{kkz,kkz2} was in a sense of dualities not complete and only one facet of this brilliant diamond. This facet is shared by two and three color cases. There was only one duality and in the case of three colors symmetric was only one section of the phase diagram, namely, $(\nu,\nu_{5})$ and it was hard to imagine different sections and that sections were not such symmetric. 
But in the two color case the phase diagram is extremely symmetric and one can easily imagine all the sections and they are very similar (have the same structure and can be mapped between each other by three dual symmetries). It is symmetric to such a degree (to tell the truth it has been only realized after the phase portrait was explored numerically) 
that the whole phase structure of dense quark matter with isospin and chiral imbalances in the two color case can be easily obtained and comprehended from the part of it (bearing in mind the connection with three color case, from the results of three color NJL model (without considering diquark channel)).
So the results concerning diquark condensation in two color case can be obtained just by considerations of three color case without any diquark condensation, 
which is quite interesting in itself. So one has enormously symmetric phase structure of dense quark matter with isospin and chiral imbalances in the case of two color QCD one facet of which was explored and noticed in the three color case investigations.

Our analysis performed in the framework of two-color NJL model shows that there is a very rich phase structure of dense quark matter with
isospin and chiral isospin imbalances. And there could occur a lot of phase transitions that can be interesting in various applications. For
example, at high baryon density, e.g., in neutron star or intermediate energy heavy ion collision experiment, etc, several first order phase
transitions can occur with rather moderate change of parameters. It is clear from Fig. 5 where behaviors of the order parameters (i.e. gaps or
condensates) are shown.

We performed our study mainly in the chiral limit, $m_0=0$, when the general phase portrait of the model has three dual symmetries ${\cal D}_{1,2,3}$ (\ref{IV.55})-(\ref{IV.60}). It is worth noting that when $m_0\ne 0$, only the ${\cal D}_{1}$ (\ref{IV.55}) is preserved. In this case, in our opinion, the other two dual symmetries ${\cal D}_{2,3}$ of the phase portrait of two-color quark matter are fulfilled
only approximately, and then in the region where $\mu_B,\mu_I,\mu_{I5}>m_\pi$. We draw such a conclusion based on an analysis of the duality at $m_0\ne 0$ between the CSB and charged PC phenomena carried out in Ref. \cite{kkz2}.

Note that our study confirms the previously known fact that chiral asymmetry induces the phenomenon of condensation of charged pions in a dense baryonic medium \cite{kkz1+1}. In addition, recall that it was previously shown in Ref. \cite{abuki} that in an electrically neutral medium (without chiral asymmetry) a charged PC phase is impossible. Therefore, in the future it would be of course interesting to consider the phase structure of the two-color and chirally asymmetric NJL model (1)-(2) taking into account other real parameters characterizing a dense baryon medium in neutron stars. Namely, to take into account a strong external magnetic field, as well as the fact that in the cores of compact stars, dense matter is electrically neutral and in beta equilibrium (for this it is necessary to introduce electrons into consideration) and see their effect on the charged PC phase. Of course, in this case, in the framework of the two-color QCD the results will depend on how the charges of the quarks are selected (see above comments in the text). At $N_c=3$, some information on the investigations in this direction is already available \cite{kkz4}. So we can conclude that in an electrically neutral medium in beta equilibrium, the chemical potential $\mu_{I5}$ is able to induce a charged PC phase. However, the range of parameters at which this phenomenon can be realized is much wider if both $\mu_{I5}$ and $\mu_{5}$ are nonzero (the details of the investigation have been presented in a recent paper \cite{kkz4}). Although, probably consideration of these conditions is slightly more pertinent in the case $N_c=3$, it could be useful to generalize this study to chirally asymmetric two-color quark matter as well.

In the same way, it would be interesting to note that we have also some preliminary information that the inclusion of one more chiral chemical potential $\mu_5$ into account does not spoil the dual properties of the model (1)-(2), but these dualities can be nicely demonstrated in the case of zero $\mu_5$ without considering the most general case and . Otherwise in the most general case there are a lot of parameters and there would be a lot of figures that are rather hard to track.  Of course, the phase structure becomes a more complicated, but $\mu_5$ does not break the dualities between CSB, charged PC and BSF phases of two-colored quark matter. But these dualities can be nicely demonstrated in the case of zero $\mu_5$ without additional complications. Otherwise in the most general case there are a lot of parameters and there would be a lot of figures that are rather hard to track. The influence of $\mu_5$ on different phases will be considered in a forthcoming studies. Moreover, we are going to consider the properties of dense baryonic matter with isospin and chiral asymmetries in the framework of a more realistic three-color NJL model with $\bar qq$ and diquark $qq$ channels.

\section{Acknowledgments}
R.N.Z. is grateful for support of Russian Science
Foundation under the grant \textnumero 19-72-00077 and the Foundation for the Advancement of Theoretical Physics and Mathematics BASIS grant.

\appendix

\section{Traces of operators and their products}
\label{ApB}

Let $\hat A,\hat B,...$ are some operators in the Hilbert space
$\mathbf H$ of functions $f(x)$ depending on four real variables,
$x\equiv (x^0,x^1,x^2,x^3)$. In the coordinate representation their matrix
elements are $A(x,y), B(x,y) ,...$, correspondingly, so that
 $$(\hat A f)(x)\equiv \int d^4yA(x,y)f(y),~~(\hat A\cdot \hat
 B)(x,y)\equiv \int
 d^4zA(x,z)B(z,y), ~~\mbox{etc}$$
 By definition,
 \begin{eqnarray}
{\rm Tr}\hat A\equiv\int d^4xA(x,x),~~{\rm Tr}(\hat A\cdot\hat
B)\equiv\int d^4xd^4yA(x,y)B(y,x),~~\mbox{etc}.
\label{B0}
\end{eqnarray}
Now suppose that $A(x,y)\equiv A(x-y)$, $B(x,y)\equiv B(x-y)$, i.e.
that $\hat A,\hat B$ are translationally invariant operators. Then
introducing the Fourier transformations of their matrix elements,
i.e.
\begin{eqnarray}
\overline{A}(p)=\int d^4z
A(z)e^{ipz},
~~~~~~~ A(z)=\int\frac{d^4p}{(2\pi)^4}
\overline{A}(p)e^{-ipz},~~~\mbox{etc},
\label{B3}
\end{eqnarray}
where $z=x-y$, it is possible to obtain from the above formulae
 \begin{eqnarray}
{\rm Tr}\hat A=A(0)\int d^4x=\int\frac{d^4p}{(2\pi)^4}
\overline{A}(p)\int d^4x.
\label{B4}
\end{eqnarray}
If there is an operator function $F(\hat A)$, where $\hat A$ is a
translationally invariant operator, then in the coordinate
representation its matrix elements depend on the difference $(x-y)$.
Obviously, it is possible to define the Fourier transformations
$\overline{F(A)}(p)$ of its matrix elements, and the following
relations are valid ($\overline{A}(p)$ is the Fourier transformation (\ref{B3})
for the matrix element $A(x-y)$):
\begin{eqnarray}
\overline{F(A)}(p)=F(\overline{A}(p));~~~
{\rm Tr}F(\hat A)=\int\frac{d^4p}{(2\pi)^4}
F(\overline{A}(p))\int d^4x.
\label{B5}
\end{eqnarray}
Finally, suppose that $\hat A$ is an operator in some internal
$n$-dimensional vector space, in addition. Evidently, the same is
valid for the Fourier transformation $\overline{A}(p)$ which is now
some $n\times n$ matrix. Let $\lambda_i(p)$ are eigenvalues of the
$n\times n$ matrix $\overline{A}(p)$, where  $i=1,2,..,n$. Then
\begin{eqnarray}
{\rm Tr}F(\hat A)=\int\frac{d^4p}{(2\pi)^4}
\mbox{tr}F(\overline{A}(p))\int
d^4x=\sum_{i=1}^{n}\int\frac{d^4p}{(2\pi)^4}
F(\lambda_i(p))\int d^4x.
\label{B6}
\end{eqnarray}
In this formula we use the notation tr for the trace of any operator
in the internal $n$-dimensional vector space only, whereas the
symbol Tr means the trace of an operator both in the coordinate and
internal spaces. In particular, if $F(\hat A)=\ln(\hat A)$, then it follows from (\ref{B6}) that (here we use a well-known relation
$\ln\det (\hat A)={\rm Tr}\ln (\hat A)$)
\begin{eqnarray}
\ln\det (\hat A)={\rm Tr}\ln (\hat A)=\sum_{i=1}^{n}\int\frac{d^4p}{(2\pi)^4}
\ln(\lambda_i(p))\int d^4x=
\int\frac{d^4p}{(2\pi)^4}
\ln(\det\overline{A}(p))\int d^4x.
\label{B7}
\end{eqnarray}

\section{Dual invariance of the four-fermion Lagrangian (\ref{IV.100})}
\label{ApA}

Let us show that the dual symmetry ${\cal D}_1$ (see in Eq. (\ref{IV.55})) of the thermodynamic potential is based on the invariance of the
four-fermion Lagrangian (\ref{IV.100}) with respect to some discrete transformation of both spinor fields and model parameters
(coupling constants, chemical potentials, etc.)
Here we suppose that bare quark mass $m_0$ is not zero. In this case it is convenient to divide four-fermion contribution to this Lagrangian into
four terms,
\begin{eqnarray}
\big (4F\big )_{csb1}&=&(\bar qq)^2\sim\sigma^2,~~~\big (4F\big )_{csb2}=(\bar qi\gamma^5\tau_3 q)^2\sim\pi_0^2, \label{A1}\\
\big (4F\big )_{pc}&=&\Big [(\bar qi\gamma^5\tau_1 q)^2+(\bar qi\gamma^5\tau_2 q)^2\Big]\sim \pi_1^2+\pi_2^2, \label{A2}\\
\big (4F\big )_{bsf}&=&\Big [\big (\bar qi\gamma^5\sigma_2\tau_2q^c\big )\big (\overline{q^c}i\gamma^5\sigma_2\tau_2 q\big )\Big]\sim |\Delta|^2
\equiv \Delta^*\Delta, \label{A3}
\end{eqnarray}
where the auxiliary bosonic fields $\sigma, \vec\pi$ etc are introduced in Eq. (\ref{IV.3}). In the notation of quark flavors, $q_u$ and $q_d$,
these quantities look like
\begin{eqnarray}
\big (4F\big )_{csb1}&=&\Big [\bar q_uq_u+\bar q_dq_d\Big]^2,~~~\big (4F\big )_{csb2}=-\Big [\bar q_u\gamma^5q_u-\bar q_d\gamma^5q_d\Big]^2, \label{A4}\\
\big (4F\big )_{pc}&=&-\Big [\bar q_u\gamma^5q_d+\bar q_d\gamma^5q_u\Big]^2+\Big [\bar q_u\gamma^5q_d-\bar q_d\gamma^5q_u\Big]^2=-4
(\bar q_u\gamma^5q_d)(\bar q_d\gamma^5q_u), \label{A5}\\
\big (4F\big )_{bsf}&=&\Big [\bar q_u\sigma_2\gamma^5q_d^c-\bar q_d\sigma_2\gamma^5q_u^c\Big]\Big [\overline{ q_u^c}\sigma_2\gamma^5q_d-
\overline{q_d^c}\sigma_2\gamma^5q_u\Big]=-4\big(\bar q_u\sigma_2\gamma^5q_d^c\big)\big(\overline{q_d^c}\sigma_2\gamma^5q_u\big). \label{A6}
\end{eqnarray}
To obtain the last expression in Eq. (\ref{A6}), we use in squarte brackets there two rather general relations,
\begin{eqnarray}
\overline{\chi^c}U \eta&=&-\overline{\eta^c}\big(CUC\big)^T \chi,~~\bar\psi V \zeta^c=-\bar\zeta\big(CVC\big)^T \psi^c, \label{A7}
\end{eqnarray}
in which $\chi,\eta,\psi$ and $\zeta$ are Dirac spinors, $U$ and $V$ are matrices in spinor space (for other notations see the explanations
to Eq. (\ref{IV.1})).

Now, it  is clear that under the transformation $q_d\rightarrow \sigma_2q_d^c$ (the quark field $q_u$ does not changed) we have
$\sigma_2q_d^c\rightarrow -q_d$. In this case  $\big (4F\big )_{pc}\leftrightarrow\big (4F\big )_{bsf}$, as it can easily be seen from Eqs.
(\ref{A5}) and (\ref{A6}).
Moreover, since $\overline{(\sigma_2q_d^c)}\sigma_2q_d^c=\bar q_dq_d$ and $\overline{(\sigma_2q_d^c)}\gamma^5 \sigma_2q_d^c=
\bar q_d\gamma^5q_d$, the structures
$\big (4F\big )_{csb1}$ and $\big (4F\big )_{csb2}$ in Eq. (\ref{A4}) remain intact under this transformation. It is also clear that in this
case the quark number and isospin densities
pass into each other, i.e. $\bar q\gamma^0 q \leftrightarrow\bar q\gamma^0 \tau_3 q$, but the free Dirac Lagrangian
$\bar q \big (i\hat\partial-m_0\big )q$ and chiral isospin density $\bar q\gamma^0\gamma^5
\tau_3 q$ are invariant. As a result, we see that under the following simultaneous discrete transformations of the spinor fields and
chemical potentials,
\begin{eqnarray}
\widetilde{\cal D}_1:~~q_d\longrightarrow \sigma_2q_d^c,~~\mu_B\longleftrightarrow \mu_I,\label{A8}
\end{eqnarray}
the Lagrangian (\ref{IV.100}) is invariant. 
Just this symmetry leads to the invariance of the mean-field approximation (\ref{IV.73}) as well as of the whole TDP of the model
with respect to the dual tranformation ${\cal D}_1$ (\ref{IV.55}) and, hence, to the dual correspondence
between charged pion and diquark condensation phenomena.

\section{Calculation of the TDP (\ref{IV.73}) at some particular values of condensates}
\label{ApC}
{\bf The case $\Delta=0$.} To obtain the expression for the TDP (\ref{IV.73}) in this case, we first look at the Eqs.
(\ref{IV.18})-(\ref{IV.22}) from which it follows that at $\Delta=0$ and $N_c=2$
\begin{eqnarray}
&&\hspace{-1cm}{\cal S}_{\rm
{eff}}(M,\pi_1,|\Delta|=0)
=-\int d^4x\frac{(M-m_0)^2+\pi^2_1}{4H}\nonumber\\&-&
i\ln\mbox {det}\Big [\big (-i\hat\partial-M-\widetilde{\cal M}\gamma^0  +i\gamma^5\tau_1\pi_1\big )
\big (i\hat\partial-M-\gamma^0{\cal M} -i\gamma^5\tau_1\pi_1\big )\Big ]\label{IV.180}
\end{eqnarray}
(note that in the square brackets in Eq. (\ref{IV.180}) we have operators which do not act in the color space). Then, using for the operators
the momentum space representation, we get (as in the similar Eq. (\ref{IV.32}))
\begin{eqnarray}
{\cal S}_{\rm
{eff}}(M,\pi_1,|\Delta|=0)
=-\int d^4x\frac{(M-m_0)^2+\pi^2_1}{4H}
-i\int\frac{d^4p}{(2\pi)^4}\ln\Big [\det\big(\overline D_1(p)\big)\det\big(\overline D_2(p)\big)\Big ]\int d^4x,\label{IV.181}
\end{eqnarray}
where $\overline D_1(p)=\big(-\hat p-M-\widetilde{\cal M}\gamma^0  +i\gamma^5\tau_1\pi_1\big)$ and $\overline D_1(p)=
\big(\hat p-M-\gamma^0{\cal M} -i\gamma^5\tau_1\pi_1\big)$ are the 8$\times$8 matrices in the direct production of the 4-dimensional spinor
and 2-dimensional flavor spaces. Their determinants are the following,
\begin{eqnarray}
\det\big(\overline D_1(p)\big)=P_4^-(\eta)P_4^+(\eta),~~\det\big(\overline D_2(p)\big)=P_4^-(\tilde\eta)P_4^+(\tilde\eta),\label{IV.182}
\end{eqnarray}
where $\eta=p_0+\mu$, $\tilde\eta=p_0-\mu$ and $P_4^\pm(x)$ are the 4-th order polynomials,
$P_4^\pm(x)=x^4-2ax^2\pm bx+c$, where $|\vec p|=\sqrt{p_1^2+p_2^2+p_3^2}$ and
\begin{eqnarray}
a=M^2+\pi_1^2+|\vec p|^2+\nu^2+\nu_{5}^2;~~b=8|\vec p|\nu\nu_{5};~~
c=a^2-4|\vec p|^2(\nu^2+\nu_5^2)-4M^2\nu^2-4\pi_1^2\nu_5^2-4\nu^2\nu_5^2.\label{IV.183}
\end{eqnarray}
In this case, using the definition (\ref{IV.19}) of the TDP, it is possible to get from Eq. (\ref{IV.181}) the following expression for the
TDP (\ref{IV.73}) at $\Delta=0$,
\begin{eqnarray}
&&\Omega(M,\pi_1,|\Delta|=0)
=\frac{(M-m_0)^2+\pi^2_1}{4H}
+i\int\frac{d^4p}{(2\pi)^4}\ln\Big [P_4^-(\eta)P_4^+(\tilde\eta)P_4^+(\eta)P_4^-(\tilde\eta)\Big ].\label{IV.184}
\end{eqnarray}
In Ref. \cite{kkz} the phase structure of the NJL model without diquark channel was investigated in the mean-field approximation
(or in the leading order of the large-$N_c$ expansion) and with the same chemical potential content, i.e. at
$\mu_B\ne 0,\mu_I\ne 0,\mu_{I5}\ne 0$. There only the single $(\bar q q)$ channel with coupling constant $G$ was taken into account and quarks
are considered as three color. It is possible to show that at $2H=G$ the TDP (\ref{IV.184}) is actually
the doubled TDP obtained in Ref. \cite{kkz}. It can be explained by the following.
If one divides the coupling constant $G$ by the number of colors $N_{c}$, it can be shown that the TDP divided by $N_{c}$ does not depend
on the number of colors if one keep $G$ the same for all number of colors. In the present paper neither $H$ nor the model TDP (\ref{IV.19})
is divided by $N_{c}$. That is why the TDP (\ref{IV.184}) in our case is equal to doubled TDP 
in three color case obtained in Ref. \cite{kkz} if one substitute $G=2H$. Now let us discuss the fit of the model parameters. The fit in the
two color case can be constructed from the fit in three color case by the following $N_{c}$ scaling of physical quantities
$f_{\pi}\sim \sqrt{N_{c}}$ and $\langle \bar{\psi} \psi \rangle \sim N_{c}$  \cite{brauner1}. It can be easily shown that it is equivalent to
the scaling of coupling constant $G\sim 1/N_{c}$. Together with the above statement that TDP divided by $N_{c}$ does not depend on the number
of colors, if $G$ is divided by $N_{c}$ (scaling $G\sim 1/N_{c}$), one can see that the phase structure of NJL model with quark-antiquark
channel should not depend on the number of colors. If one divide the coupling constant $G$ by the number of colors $N_{c}$ as it was done, for example, in \cite{kkz} then the TDP of the system divided by $N_{c}$ does not depend on the number of colors.


Hence, if diquark condensate $\Delta$ is equal to zero in the
framework of two-color NJL model, then the phase structures of this model and the ordinary three-color NJL model coincide at $2H=G$ and the
same values of the cutoff parameter $\Lambda$.
Note that at $\Delta=0$ the quantities (products) $\lambda_1(p)\lambda_2(p)$ and $\lambda_3(p)\lambda_4(p)$ from the expression (\ref{IV.73})
are just the products $P_4^-(\eta)P_4^+(\tilde\eta)$ and $P_4^+(\eta)P_4^-(\tilde\eta)$, respectively. Moreover, the results presented after
Eq. (\ref{IV.181}) can be obtained with the help of any program of analytical calculations. Let us denote by $x_1,...,x_4$ the roots of the
polynomial $P_4^-(x)$. Then it is clear that the polynomial $P_4^+(x)$ has the same roots but with opposite sign, $-x_i$ ($i=1,...,4$). So,
\begin{eqnarray}
&&P_4^-(x)P_4^+(x)=(x^2-x_1^2)(x^2-x_2^2)(x^2-x_3^2)(x^2-x_4^2)\label{IV.185}
\end{eqnarray}
and (recall that below $\eta,\tilde\eta=p_0\pm\mu$)
\begin{eqnarray}
&&\Omega(M,\pi_1,|\Delta|=0)
=\frac{(M-m_0)^2+\pi^2_1}{4H}
+i\int\frac{d^4p}{(2\pi)^4}\ln\Big [(\eta^2-x_1^2)(\eta^2-x_2^2)(\eta^2-x_3^2)(\eta^2-x_4^2)\nonumber\\
&&~~~~~~~~~~~~~~~~~~~~~~(\tilde\eta^2-x_1^2)
(\tilde\eta^2-x_2^2)(\tilde\eta^2-x_3^2)(\tilde\eta^2-x_4^2)\Big ].
\label{IV.186}
\end{eqnarray}
Integrating in Eq. (\ref{IV.186}) over $p_0$ (taking into account Eq. (\ref{IV.39})), we have
\begin{eqnarray}
&&\Omega(M,\pi_1,|\Delta|=0)
=\frac{(M-m_0)^2+\pi^2_1}{4H}
-\sum_{i=1}^4\int\frac{d^3p}{(2\pi)^3}\Big [|\mu-x_i|+|\mu+x_i|\Big ].
\label{IV.187}
\end{eqnarray}
Taking into account in the square brackets of Eq. (\ref{IV.187}) only the first term, $|\mu-x_i|$, one can obtain the contribution of the
eigenvalues $\lambda_1(p)$ and $\lambda_2(p)$ to expression (\ref{IV.73}) for the TDP. whereas the second term there is responsible for the
contribution of the eigenvalues $\lambda_3(p)$ and $\lambda_4(p)$ to the TDP (\ref{IV.73}).

{\bf The case $M\ne 0$, $\pi_1=0$,  $\Delta=0$}. In this case the roots $x_i$ of the polynomial $P_4^-(x)$ looks like \cite{kkz}
\begin{eqnarray}
&&x_{1,2}=-\nu\pm\sqrt{M^2+(|\vec p|-\nu_5)^2},~~x_{3,4}=\nu\pm\sqrt{M^2+(|\vec p|+\nu_5)^2}. \label{IV.188}
\end{eqnarray}
Hence,  it follows from Eq. (\ref{IV.187}) that $\Omega(M,\pi_1=0,|\Delta|=0)$ at $m_0=0$ is equal to the projection $F_1(M)$ (\ref{IV.189}).

{\bf The case $M=0$, $\pi_1\ne 0$, $\Delta=0$}. To obtain the TDP in this case, one can take into account that at $m_0=0$ it is invariant under the dual
transformation ${\cal D}_3$ (\ref{IV.60}). So, in this case the TDP is equal to the projection
$F_1(M)$ (\ref{IV.189}), in which we should perform the changes $M\to\pi_1$ and $\nu\leftrightarrow\nu_5$. The result is the expression
$F_2(\pi_1)$ (\ref{IV.190}).

{\bf The case $M=0$, $\pi_1= 0$, $\Delta\ne 0$}. Finally, to get the TDP (\ref{IV.73}) in this particular case, one can simply take into
account that at $m_0=0$ it is invariant under the dual transformation ${\cal D}_2$ (\ref{IV.60}). So, in this case it is no more than the
expression (\ref{IV.189}), in which $M\to |\Delta|$ and $\nu_5\leftrightarrow\mu$ changes should be performed. The result is the expression
$F_3(|\Delta|)$ (\ref{IV.191}).

Note that in all projections $F_i$  of the TDP on the condensate axes, the first two terms in square brackets
of the expressions (\ref{IV.189})-(\ref{IV.191}) correspond to the contribution of the eigenvalues $\lambda_1(p)$ and $\lambda_2(p)$ to the
expression (\ref{IV.73}) of the TDP. However, the last two terms in square brackets there correspond to the contribution of $\lambda_3(p)$ and $\lambda_4(p)$ to it.

\end{document}